\documentclass[aps,pre,superscriptaddress,amsmath]{revtex4-1}
\usepackage{amssymb}
\usepackage{amsmath}
\usepackage{graphicx}
\usepackage{amsbsy}
\usepackage{xr}
\usepackage{float}
\usepackage{bm}
\usepackage{color}
\usepackage{gensymb}
\usepackage{caption}
\usepackage{subcaption}
\usepackage[space]{grffile}
\definecolor{cream}{RGB}{222,217,201}
\usepackage{siunitx}
\DeclareSIUnit[number-unit-product = {\,}]
\cal{cal}
\DeclareSIUnit\kcal{\kilo\cal}
\DeclareSIUnit\kcal{\kilo\joule\per\mole}
\DeclareSIUnit\molar{\mole\per\cubic\deci\metre}
\DeclareSIUnit\Molar{\textsc{m}}
\usepackage{multirow}

\begin{document}

\title{An integrated DEM-FEM approach to study breakage in packing of glass cartridges on a conveyor belt}

\author{Daniela Paola Boso}
\email{daniela.boso@unipd.it}
\affiliation{Dipartimento di Ingegneria Civile ed Ambientale,
	Universit\'a degli Studi di Padova,
	via Marzolo 9, 35131 Padova, Italy}
\author{Tommaso Braga}
\affiliation{Dipartimento di Scienze Molecolari e Nanosistemi, 
Universit\`{a} Ca' Foscari di Venezia
Campus Scientifico, Edificio Alfa,
via Torino 155,30170 Venezia Mestre, Italy}
\author{Simone Ravasini}
\affiliation{Dipartimento di Ingegneria Civile ed Ambientale,
	Universit\'a degli Studi di Padova,
	via Marzolo 9, 35131 Padova, Italy}
\author{Tatjana \v{S}krbi\'{c}}
\affiliation{Dipartimento di Scienze Molecolari e Nanosistemi, 
Universit\`{a} Ca' Foscari di Venezia
Campus Scientifico, Edificio Alfa,
via Torino 155,30170 Venezia Mestre, Italy}
\author{Andrea Puglisi}
\affiliation{Istituto dei Sistemi Complessi CNR, Dipartimento di Fisica, Universit\'a ``La Sapienza''
di Roma, p.le A. Moro 2, I-00185 Roma, Italy}
\author{Odra Pinato}
\affiliation{SG Lab Analytics, Nuova Ompi, Stevanato Group
  Via Molinella, 17, 35017 Piombino Dese Padova  Italy}
\author{Alberto Chillon}
\affiliation{SG Lab Analytics, Nuova Ompi, Stevanato Group
  Via Molinella, 17, 35017 Piombino Dese Padova  Italy}
\author{Maria Chiara Frare}
\affiliation{SG Lab Analytics, Nuova Ompi, Stevanato Group
  Via Molinella, 17, 35017 Piombino Dese Padova  Italy}
\author{Achille Giacometti}
\email{achille.giacometti@unive.it}
\affiliation{Dipartimento di Scienze Molecolari e Nanosistemi, 
Universit\`{a} Ca' Foscari di Venezia
Campus Scientifico, Edificio Alfa,
via Torino 155,30170 Venezia Mestre, Italy}

\date{\today}

\begin{abstract}
The use of glass for pharmaceutical new applications such as high-technology drugs, requires the strictest container inertness.  
A common theme of paramount importance in glass container integrity preservation is the detailed mechanism driving the sudden failure due the crack propagation. Using a combination of Discrete Element Method (DEM) and Finite Element Method (FEM), a stress map for glass cartridges packed into an accumulation table and transported by a conveyor belt at a fixed velocity is obtained under realistic conditions. The DEM calculation provides a full description of the dynamics of the cartridges, as approximated by an equivalent sphere, as well as the statistics of the multiple collisions. The FEM calculation exploits this input to provide the maximum principal stress of different pairs as a function of time. Our analysis shows that, during their transportation on the conveyor belt, the cartridges are subject to several shocks of varying intensities. Under these conditions, a crack may originate inside the cartridge in the area of maximal tensile stress, and propagate outward. Estimated stresses are found in good agreement with real systems. 
\end{abstract}
\maketitle
\section{Introduction}
\label{sec:introduction}
Frictional contacts between macroscopic objects may lead to deformation and breakage. The analysis of conditions for breakage in a large packing of objects with multiple contacts and collisions requires the evaluation of phenomena at different scales, from large scale force propagation (with strong fluctuations in space and time) down to local microscale stress accumulation. Here we analyse a case study, inspired by an industrial problem, which reveals several unexpected facets and illustrates the connection between material mechanics, physics and granular statistical physics.

\begin{figure}[h]
  \includegraphics[width=0.55\textwidth]{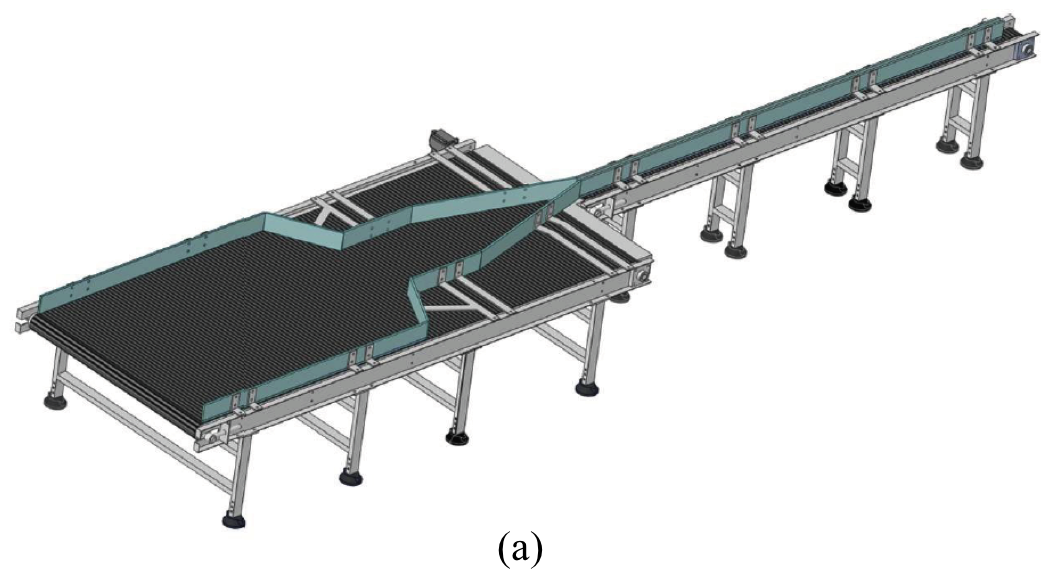}
  \includegraphics[width=0.55\textwidth]{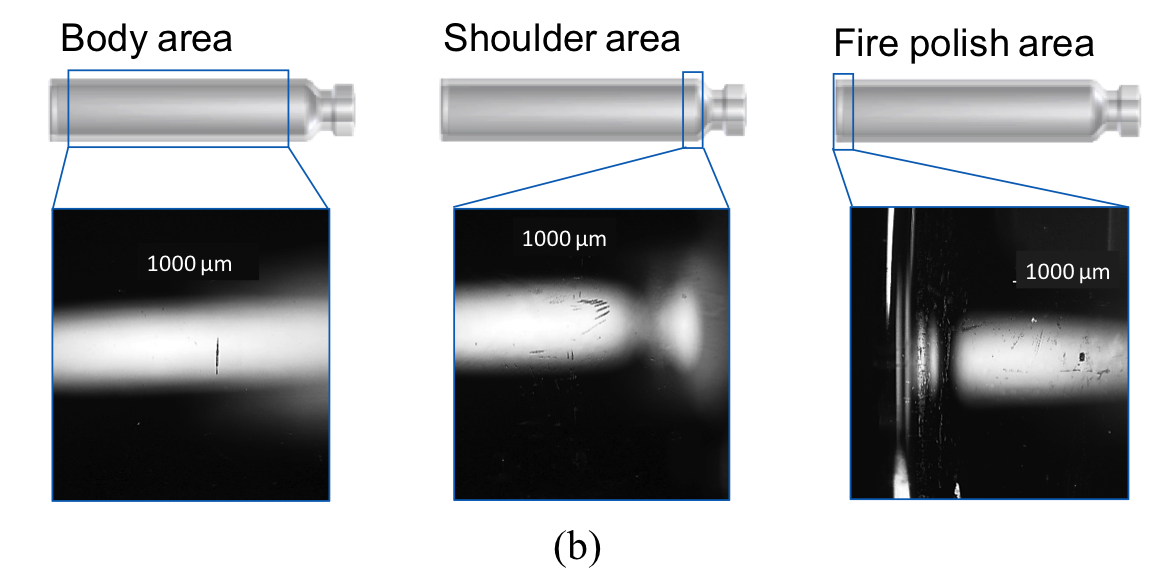} 
\caption{(a) A typical production line with the accumulation table for the cartridge; (b) Three cartridges with characteristic cracks in different parts of the cartridge: body area (left), shoulder area (center), fire polish area (right)}
\label{fig:fig1}       
\end{figure}

The use of glass for pharmaceutical keeps increasing, especially for new applications such as high-technology drugs, requiring the strictest container inertness. Indeed, high-end biotech products keep entering the market, posing new technical and regulatory challenges for the development of injection systems \cite{Schaut17}. Among the primary containers, syringes and cartridges are becoming the key players thanks to their superior performances and easy usability, replacing the reconstitution filling procedures for vials. Following the market trends, new manufacturing processes and advanced technologies have been developed, making available a new range of primary packaging aimed at increasing patient safety and reducing the risk of market recall. Regarding this, one of the most critical topics remains the mechanical resistance: glass breakages can lead to serious issues like container integrity failure, microbiological contamination of a sterile package and glass particulates generation. Almost always, the decrease of the mechanical resistance \cite{Mould67} is caused by strength-reducing damage introduced during forming and handling processes. For this reason studies for risk assessment, based on physical and functional factors, become fundamental, starting from the early stages of the primary container manufacturing. The above problem is not confined to technological issues but it also bears important consequences in terms of slow dynamics non-linearity in brittle materials \cite{Yoritomo19}, and in the problem of slow flows of granular materials drained by conveyor belts \cite{Janda15,Perng06}, not dissimilar from the widespread problem of granular discharge \cite{Sheldon10}. From this perspective, a numerical modelling that simulates the critical steps of the container life-cycle can be a powerful tool to reduce the risk of those adverse events. Figure \ref{fig:fig1} shows the typical conditions that our calculation would aim at reproducing.

\begin{figure}[h]
  \includegraphics[width=0.35\textwidth]{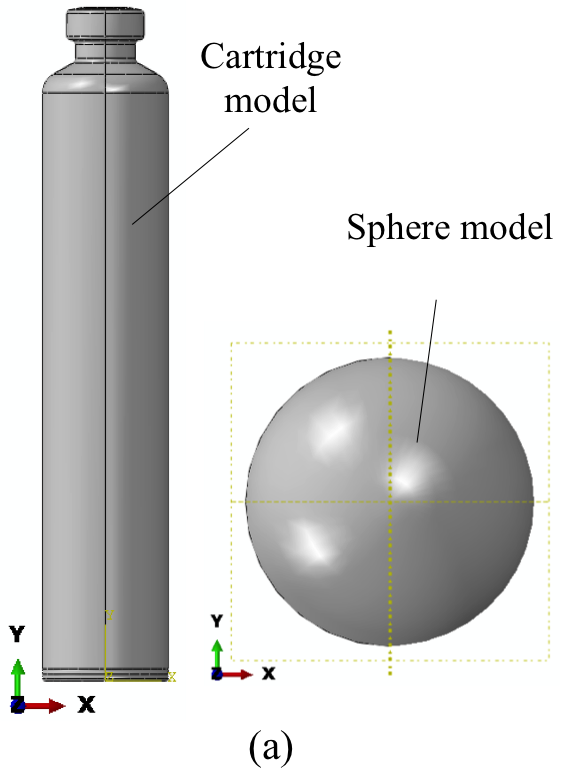}
  \includegraphics[width=0.35\textwidth]{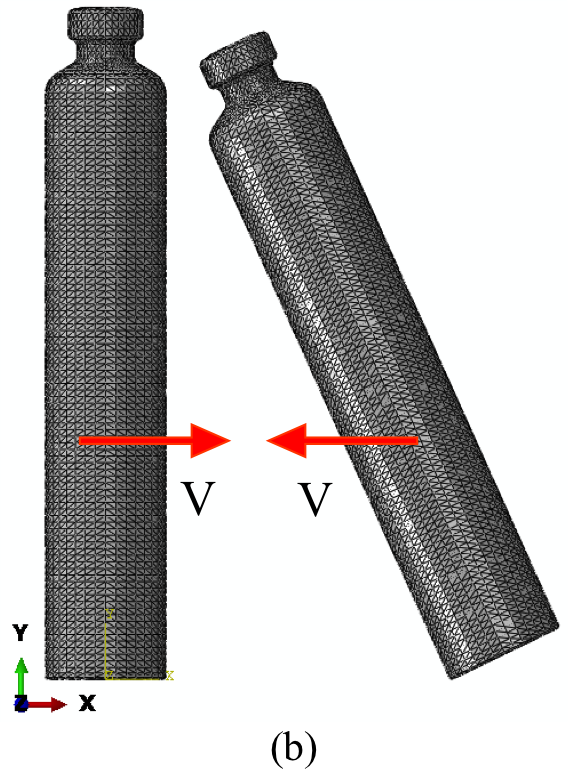}
  \caption{(a) Effective spheres representing the cartridge in DEM simulations; (b) The collision geometry of two cartridges. 
  \label{fig:fig2}}
\end{figure}

In this paper, we will discuss the glass breakage issue for cylindrical cartridges, that are one of the most commonly used containers in pharmaceutical processes.
Cartridges are directed through an accumulation table toward the filling line (see Fig. \ref{fig:fig1}(a)). During this process, they experience a high packing condition that may lead to the onset of invisible micro-fractures (examples are given in Fig.\ref{fig:fig1}(b)) that affect the glass integrity and their safe use in the filling process. The idea of the present study is to identify the critical conditions for this to occur. This will be attained by a three-step process.
\begin{description}
  \item[1.] Discrete Element Method (DEM) and Finite Element Method (FEM) to justify and to define the equivalent sphere model together with its parameters (input data for step 2) \\
Firstly, the optimal geometrical and dynamical conditions will be studied using a Discrete Element Method (DEM) calculation \cite{Poschel10} as provided by the Yade package \cite{Smilauer15}. To this aim, we have built a model for a cylindrical cartridge as shown in Figure S1 of Supplementary Information. Each cylindrical cartridge is composed by 632 identical small beads -- a smaller number of small sphere would imply a too poor resolution and make the shape of the simulated cartridges unfaithful to the real ones.
  A full analysis of the dynamics of many such cartridges proves however a daunting task, especially under realistic conditions, because it would involve the bookkeeping of each of the sphere composing each single cartridge, with a corresponding huge computational cost. Therefore we provide evidence that under high packing conditions the most relevant features of the dynamics can be still captured by using a set of equivalent spheres, that were obtained by the same glass material and total mass of the cartridges. A visual representation of this is displayed in Figure \ref{fig:fig2}(a). As cartridges in the filling line are highly packed, their dynamics is essentially driven by by translations in the plane perpendicular to gravity and rotations (for instance due to friction) around the axis of symmetry of the cylindrical cartridge, i.e. other movements such as inclinations/falling of a cartridge are negligible.
  Hence, the use of an equivalent sphere turns out to be fully justified at this stage, as it will be further discussed below. A large number of these equivalent spheres is then inserted into a computational box mimicking the accumulation table displayed in Figure \ref{fig:fig1} (a), with an underline conveyor belt whose velocity can be tuned. In this way, different geometries matching those of the accumulation table can also be tested, and the final optimal configuration quickly identified. See Section \ref{subsec:comparison}
\item[2.] DEM simulations to obtain input data for step 3;\\
  In the second step, we will perform a direct DEM calculation of the velocity field of the cartridge population (as represented by their equivalent spheres), as well as the number of contact points for selected items, as a function of time, during the process of accumulation toward the filling line. The velocity field and other dynamical information derived from the DEM analysis will be then used as an input to perform a Finite Element Method (FEM) analysis \cite{Johnson12} on a small cluster of cartridges, using the Abaqus package \cite{Abaqus}. See Section \ref{subsec:scaling}
\item[3.] FEM simulations to assess the likelihood of crack initiation;\\
  This final step will provide a complete stress map of the cartridges, which allows the identification of the critical regions where microfractures can occur under that specific geometry. See Section \ref{subsec:stress}.
\end{description}
The plan of the paper is as follows. In Section \ref{sec:methods} we discuss the mapping of a cartridge into an equivalent sphere, and the theoretical background underlying the DEM and FEM analyses. In Section \ref{sec:results}, we report all the obtained results, first from the DEM to identify the optimal geometry of the filling line, as well as the velocity field, and then from the FEM to study the stress map of a small cluster of cartridges subject to collisions as predicted by the DEM calculation (Section \ref{sec:discussion}). Finally, Section \ref{sec:conclusions} summarizes some key messages and presents some possible future perspectives. 

\section{Theoretical background and Methods}
\label{sec:methods}
\subsection{The Hertz contact model for full spheres, Reissner theory of thin shells and the cartridge behavior}
\label{subsec:hertz}
\begin{figure}[h]
	\centerline{\includegraphics[width=0.80\textwidth]{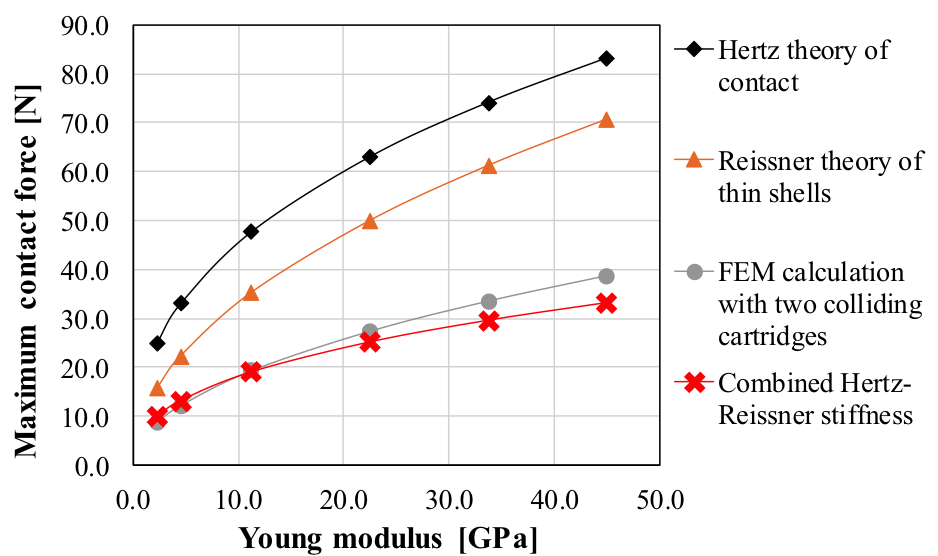}}
	\caption{Maximum contact force as computed using the finite element method, Reissner only theory, Hertz only theory and a combination of Reissner and Hertz models.
	\label{fig:fig3}}
\end{figure}

Consider a cartridge travelling on a filling line and impacting a neighboring one as schematically depicted in Figure \ref{fig:fig2}(b). The total deformation can be split into two main contributions: the mutual indentation depth and the deflection of the thin glass walls due to membrane and bending actions. If small displacements and strains along with linear elasticity hold, then the two terms can be calculated separately, and the system can be modeled by combining the corresponding stiffness in series. Furthermore, since the contact area is limited to a small zone, the two contacting bodies can be assimilated to two equivalent empty spheres, keeping the same mass of the cartridges. Under these assumptions, the maximum mutual indentation can be calculated following Hertz' theory of non-adhesive elastic contact between two spheres \cite{Johnson12}
\begin{equation}
	F=k_Hd^{\frac{3}{2}}_H
	\label{sec3:eq1}
\end{equation}
where $k_H$ is the Hertz stiffness
\begin{equation}
	k_H=\frac{4}{3}\sqrt{R^*}E^*
	\label{sec3:eq2}
\end{equation}
and:
\begin{equation}
	\frac{1}{R^*}=\frac{1}{R_1}+\frac{1}{R_2}
	\label{sec3:eq3}
\end{equation}
\begin{equation}
	\frac{1}{E^*}=\frac{(1-\nu^2_1)}{E_1}+\frac{(1-\nu^2_2)}{E_2}
	\label{sec3:eq4}
\end{equation}
The membrane and bending deformation is evaluated after Reissner theory \cite{Reissner46a,Reissner46b}, and the maximum deflection of one shell is given by the linear relation
\begin{equation}
	F=k_Rd_R
	\label{sec3:eq5}
\end{equation}
where:
\begin{equation}
	k_R=\frac{4}{\sqrt{3(1-\nu^2)}}\frac{Et^2}{R}
	\label{sec3:eq6}
\end{equation}
For a two-cartridge system accounting for both Hertz indentation and Reissner theory, the energy balance equation can be written as
\begin{equation}
	\frac{1}{2}m_1v_1^2+\frac{1}{2}m_2v_2^2=\frac{1}{2}k_{R1}d_{R1}^2+\frac{1}{2}k_{R2}d_{R2}^2+\frac{2}{5}k_Hd_H^{\frac{5}{2}}+\frac{1}{2}(m_1+m_2)v_{12}^2
	\label{sec3:eq7}
\end{equation}
The relative velocity $\mathbf{v}_{12}$ is obtained by imposing the linear momentum balance
\begin{equation}
	\mathbf{v}_{12}=\frac{m_1\mathbf{v}_1+m_2\mathbf{v}_2}{m_1+m_2}
	\label{sec3:eq8}
\end{equation}
Analogously to $E^*$ and $R^*$, we can define the reduced mass as:
\begin{equation}
	\frac{1}{m^*}=\frac{1}{m_1}+\frac{1}{m_2}
	\label{sec3:eq9}
\end{equation}
Equations (\ref{sec3:eq8}) and (\ref{sec3:eq9}) substituted in Equation (\ref{sec3:eq7}) lead to:
\begin{equation}
	m^*v^2_{rel}=k_{R1}d_{R1}^2+k_{R2}d_{R2}^2+\frac{4}{5}k_Hd_H^\frac{5}{2}
	\label{sec3:eq10}
\end{equation}
where $\mathbf{v}_{rel}$ is the relative velocity of the two objects:
\begin{equation}
	|\mathbf{v}_{rel}|=|\mathbf{v}_{1}-\mathbf{v}_{2}|
	\label{sec3:eq11}
\end{equation}
Considering (\ref{sec3:eq1}) and (\ref{sec3:eq5}), Equation (\ref{sec3:eq10}) can be written as:
\begin{equation}
	m^*v_{rel}^2=\frac{F^2}{k_{R1}}+\frac{F^2}{k_{R2}}+\frac{4}{5}\frac{F^{\frac{5}{3}}}{k_{H}^{\frac{2}{3}}}
	\label{sec3:eq12}
\end{equation}
which is an implicit expression for the maximum force transmitted, including the effect of both Hertzian and membrane-bending deformation for the two empty spheres. Equation (\ref{sec3:eq12}) can be easily solved numerically for the maximum force F. As an alternative, assuming that Hertz contact can be expresses by a linear law \cite{Young03}
\begin{equation}
	k_{linear}=\frac{F}{d_H}=\frac{4}{3}\left(\frac{15}{16}\right)^{\frac{1}{5}}R^{*\frac{2}{5}}E^{*\frac{4}{5}}m^{*\frac{1}{5}}v_{rel}^\frac{2}{5}
	\label{sec3:eq13}
\end{equation}
	an explicit expression for the contact force is obtained as:
\begin{equation}
	F=\sqrt{k_{eq}m^*}|\mathbf{v}_{rel}|
	\label{sec3:eq14}
\end{equation}
where the equivalent stiffness is simply given by:
\begin{equation}
	\frac{1}{k_{eq}}=\frac{1}{k_{R1}}+\frac{1}{k_{R2}}+\frac{1}{k_{linear}}
	\label{sec3:eq15}
\end{equation}
Figure \ref{fig:fig3} reports the contact force versus the Young modulus as computed using Hertz only theory, Reissner only theory, the approximate relation (\ref{sec3:eq15}) and a finite element analysis of the two cartridges. See Section \ref{subsec:comparison} for a discussion of the results.
\subsection{The Discrete Element Method (DEM)}
\label{subsec:discrete}
The Discrete Element Method (DEM) \cite{Poschel10} is a time-stepping algorithm that hinges upon the following calculation cycle:
\begin{enumerate}
\item Contacting forces are generated on the basis of the interactions between spheres
\item Spheres are displaced based on the forces previously computed and on the current positions
\end{enumerate}
The second step is a straightforward application of Newtons' Second Law that for a rigid body has a part related to forces
\begin{eqnarray}
\label{sec4:eq1}
m_{i} \frac{d \mathbf{v}_i}{dt} &=& \sum_{j=1}^N \left[\mathbf{F}_{ij}^{(c)}+ \mathbf{F}_{ij}^{(nc)} \right] +\mathbf{G}{i}
\end{eqnarray}
where $\mathbf{F}_{ij}^{(c)}$ and $\mathbf{F}_{ij}^{(nc)}$ are the contact and non-contact forces acting on the $i$-th sphere due to all others, and $\mathbf{G}_{i}$ is the gravity, and a part related to torques
\begin{eqnarray}
\label{sec4:eq2}
\mathbf{I}_{i} \frac{d \bm{\omega}_{i}}{dt} &=& \mathbf{M}_{i}
\end{eqnarray}
where $\mathbf{I}_{i}$ is the moment of inertia tensor of the $i$-th sphere, and $\bm{\omega}_{i}$
and $\mathbf{M}_{i}$ are the angular velocity of the $i$-th sphere and the total torque acting on the $i$-th sphere,
respectively. The Newton's equations are integrated through the
customary leapfrog scheme (also known as the Verlet scheme)~\cite{Verlet67}.

The first step, on the other hand, is a more delicate one because depends upon the model at hand. It hinges upon the force-displacement principle describing the
relationship between the relative movement of the two spheres and the contacting forces. The latter can always be decomposed in two parts: the normal direction that is always pointing from the center of one sphere to the other, and strongly depends on the contact (overlap) between the two bodies; and the tangential direction, whose relative movement is composed of a rotation and a translation components.

Once that both steps have been performed, time is advanced of one step, and the cycle is repeated again, until a prescribed number of time steps is reached.

Particular care has to be given to the boundary conditions, as the system is usually constrained by walls that may be constituted by different materials, and velocities cannot be assigned to walls.

All DEM simulations were carried out using an in-house adaptation of the Yade package \cite{Smilauer15}.

\subsection{The Finite Element Method (FEM)}
\label{subsec:finite}
Many physical problems can be posed by a set of (partial) differential equations and suitable boundary conditions. The finite element method (FEM) \cite{Zienkiewicz13} is a numerical approach to obtain an approximate solution to such boundary value problems.

In matrix notation, let $\mathbf{A}(\mathbf{u})=\mathbf{0}$ be a set of differential equations defined over a domain $\Omega$ with boundary conditions $\mathbf{B}(\mathbf{u})=\mathbf{0}$ on the boundary $\Gamma$ of the domain. In the finite element approach, the set of governing differential equations needs to be transformed into an integral form:
\begin{eqnarray}
  \label{sec5:eq1}
  \int_{\Omega} \mathbf{v}^{T} \mathbf{A}\left(\mathbf{u}\right)~d\Omega +
  \int_{\Gamma}  \overline{\mathbf{v}}^{T} \mathbf{B}\left(\mathbf{u}\right)~d\Gamma &=&0
\end{eqnarray}

where $\mathbf{v}$  and $\overline{\mathbf{v}}$  are two sets of arbitrary test functions. 
FEM cuts the spatial domain into a number of simple geometric sub-domains $\Omega_{e}$ - the finite elements - which are assumed to be interconnected only at a discrete number of common points - the nodes - on their boundaries $\Gamma_{e}$. The values $\mathbf{a}_i$ of the unknown functions $\mathbf{u}$ at the nodal points are the basic unknowns of the problem. The complete approximate solution is obtained by interpolating the nodal values  $\mathbf{a}_i$ by means of interpolation functions $\mathbf{N}_i$ (the so-called shape functions). Therefore, the finite element method seeks the solution in the following approximate form
\begin{eqnarray}
  \label{sec5:eq2}
  \mathbf{u} \approx \overline{\mathbf{u}} &=& \sum_{i}\mathbf{N}_i \mathbf{a}_i = \mathbf{N}\mathbf{a}
\end{eqnarray}

Obviously, the approximate solution (\ref{sec5:eq2}) cannot satisfy both the differential equations and the boundary conditions in a general case.

As a consequence, in Eq. (\ref{sec5:eq1}) instead of the arbitrary functions $\mathbf{v}$  and $\overline{\mathbf{v}}$ , a set of approximate functions is considered 
\begin{eqnarray}
  \label{sec5:eq3}
  \mathbf{v} &=& \sum_{j=1}^n \mathbf{w}_j \delta \mathbf{a}_j \\
  \overline{\mathbf{v}} &=& \sum_{j=1}^n \overline{\mathbf{w}}_j \delta \mathbf{a}_j
\end{eqnarray}
where $\delta \mathbf{a}_j$ are arbitrary and $n$ is the number of unknowns of the problem. In this way, the integral form (\ref{sec5:eq1}) leads to a set of ordinary algebraic equations
\begin{eqnarray}
  \label{sec5:eq4}
  \delta \mathbf{a}_{j}^{T}\left[\int_{\Omega} \mathbf{w}_j^{T} \mathbf{A}\left(\mathbf{N} \mathbf{a}\right)~d\Omega +
  \int_{\Gamma}  \overline{\mathbf{w}}_j^{T} \mathbf{B}\left(\mathbf{N} \mathbf{a}\right)~d\Gamma \right] &=&0 \qquad  j=1,\ldots,n
  \end{eqnarray}
from which the unknowns $\mathbf{a}_i$ can be determined.

Since $ \mathbf{A}\left(\mathbf{N} \mathbf{a}\right)$ and $ \mathbf{B}\left(\mathbf{N} \mathbf{a}\right)$ represent the errors due to the substitution of the approximate solution (\ref{sec5:eq2}) into the differential equations and boundary conditions, Eq. (\ref{sec5:eq4}) is a weighted integral of the residuals. According to the choice of the set of functions  $\mathbf{w}_j$ and $\overline{\mathbf{w}}_j$, different weighted residuals methods are obtained. In the finite element approach, the shape functions are used as weighting functions (Bubnov-Galerking method)
\begin{eqnarray}
  \label{sec5:eq5}
  \mathbf{w}_j &=& \mathbf{N}_j
\end{eqnarray}

Finally, by virtue of the property of definite integrals, the approximate solution can be obtained after each element contribution is assembled:
\begin{eqnarray}
  \label{sec5:eq6}
  \int_{\Omega} \left(\bullet \right) ~d\Omega +  \int_{\Gamma} \left(\bullet \right) ~d\Gamma &=&
\sum_{e=1}^m \left(\int_{\Omega_e} \left(\bullet \right) ~d\Omega_e + \int_{\Gamma_e} \left(\bullet \right) ~d\Gamma_e \right)
\end{eqnarray}
where $(\bullet)$ is a short-hand notation for an arbitrary function, and
where $\Omega_e$ is the domain of each element, $\Gamma_e$ its part of boundary and $m$ is the number of finite elements in the operated discretization. 
\section{Results}
\label{sec:results}
\subsection{Justification of the use of equivalent spheres}
\label{subsec:comparison}
As anticipated, one major difficulty of the DEM calculation is the very high computational cost required by the use of the full cartridge model shown in Fig.S1 of SI. We set up a computational box with geometries matching that of the real accumulation table for the cartridge, and inserted a large number of cartridges in a high packing conditions as displayed in Fig.S2(a) (details on the computational protocol will be given below). A full dynamic evolution of the system (see Fig. S2(b) for its evolution after few seconds) shows that cartridges are driven by the underlying belt by maintaining their upright position, in view of the high packing condition that they experience. Hence, all rotational degrees of freedom associated with the the tilt and possible fall for a cartridges are essentially ruled out under such conditions. The cartridge can then be replaced by a simpler object, such as a sphere, to infer the relative kinematics (i.e. velocities) and dynamics (i.e. forces) to first approximation.  The accuracy of this approximation will be eventually checked \textit{ a posteriori} by a comparison of the full evolution of a system of cartridges in one specific case.

In order to implement a mapping of a cartridge into an equivalent sphere, we need to ensure that the relative contact forces during the collisions of spheres as a function of the elapsed time will be in quantitative agreement with those originating from the cartridges. While simple in principle, care must be used as several possibilities arise. As presented in Section \ref{sec:methods}, a cartridge can be either considered as filled deformable sphere with a fully elastic behavior (Hertz theory) \cite{Landau59,Leroy85,Patricio04}, or as thin shallow spherical shell (Reissner model) \cite{Reissner46a,Reissner46b,Reissner59}, or as combination of the two (combined Hertz-Reissner model) \cite{Stein82,Kenner72,Koller86,Baghaei11}. In order to identify the optimal model, a numerical analysis of the collision of two spheres within the three different models described above, is compared with the collision of two cartridges obtained from a direct Finite Element (FE) numerical method. In all cases, the spheres were made of the same material as the cartridges and equivalent dimensions so to keep the same mass (see Table S1 in supplementary information) and the two spheres or cartridges were designed to collide at a relative velocity of $\SI{20}{\cm \cdot \sec^{-1}}$ as pictorially illustrated in Figure \ref{fig:fig2}(b). In the cartridge case, one of the two cartridges stands in the upright position and the other is tilted at a fixed angle, with the two moving toward each other with the same relative velocity as in the case of the spheres.

The same collision was repeated at increasing values of the elastic modulus $E$, and the maximum contact force was calculated with the different methods. The result of this study is reported in Figure \ref{fig:fig3} and shows that, to model two impacting cartridges, it is necessary to couple Hertz theory of contact with Reissner model for thin shells. Taken individually, Hertz and Reissner theories clearly overestimate the contact force, proving that for colliding cartridges neither the shell bending-membrane deformation nor local Hertzian deformation can be neglected. On the contrary, if both stiffnesses are considered in series, a good approximation is obtained with the equivalent spheres. A direct DEM calculation (not shown in Fig. \ref{fig:fig3}) matches those of the combined Hertz-Reissner model. As a consequence, the dynamics of the system can be well studied by means of the discrete element method taking into account equivalent spheres. As anticipated, the DEM calculation considering the realistic shape of the cartridges would turn out to be quite prohibitive from the computational point of view, with approximately two order of magnitude larger computational times. 

\subsection{Results for the dynamics of the spheres from DEM: choice of the optimal set-up}
\label{sec:results-spheres}
\begin{figure}[h]
	\centerline{\includegraphics[width=10cm]{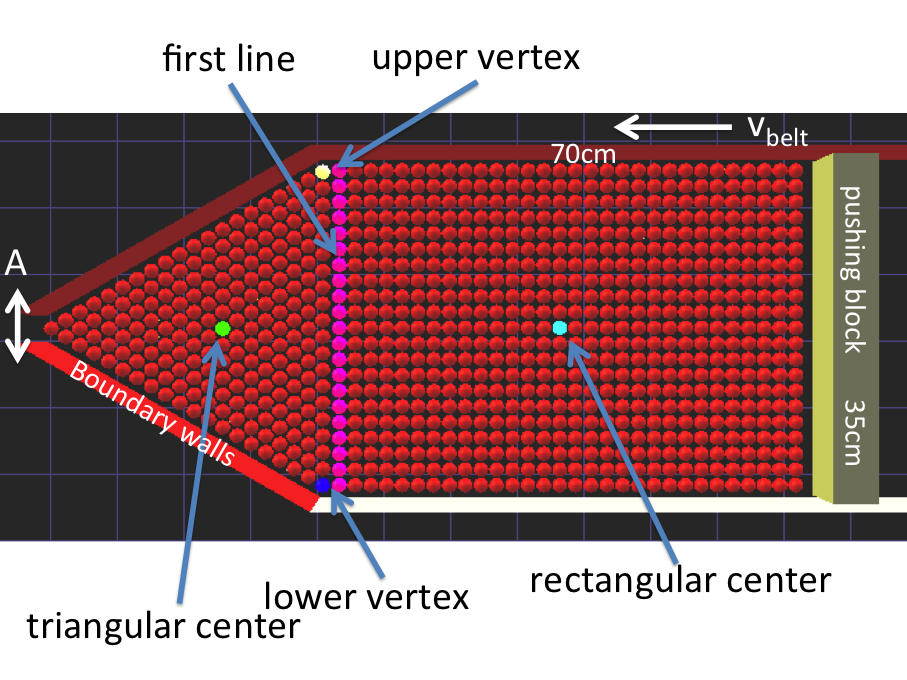}}
	\caption{Top view of our computational box with highlighted the tagged spheres used to probe the dynamics: triangular center (green), rectangular center (cyan), upper vertex (yellow), lower vertex (blue), first line (magenta). Also shown are the amplitude $A$ and the direction of the velocity $V_{\text{belt}}=\SI{20}{\cm \cdot sec^{-1}}$ for the underlying belt. 
	\label{fig:fig4}}
\end{figure}
The discrete element method will provide indications for the optimal set-up, the velocity field of the equivalent spheres as a function of time, the evolution of the number of neighboring spheres in contact with a given one, identifying the most challenging scenarios to be considered for a detailed modelling by means of finite element method.   
We first consider the dynamics of  $840$ equivalent spheres, whose calculation can be carried out under realistic values of the glass elastic modulus ($E = \SI{45}{GPa}$) within a reasonable computational time. The aim of this study is to identify the optimal set-up geometry in terms of possible contact forces and therefore stress peaks for the cartridges.

Figure \ref{fig:fig4} depicts the geometry of the computational box that represents the first part of the production line shown in Figure \ref{fig:fig1}(a). In both cases (spheres and cartridges), all entities are inserted into a planar accumulation table confined by vertical walls and whose lower plane is moving with constant belt velocity $V_{\text{belt}}$ driving them toward a constricted section ending into a small opening of amplitude $A$. The length of the tilted wall and the angle univocally define the geometry of the opening.
In the real accumulation table, an external agent is acting beyond the opening preventing a clogging of the cartridges. In order to mimic this effect,
the amplitude $A$ is subject to a harmonic oscillation of period $T$.
A number of spheres in different and strategic positions within the computational box were tagged and monitored during the dynamics, as shown in Figure \ref{fig:fig4}. We denoted as 'triangular center' (green), 'rectangular center' (cyan), 'upper vertex' (yellow), 'lower vertex' (blue) four such spheres and labelled them with different colours. In addition, we further tagged as 'first line' (magenta) all those spheres lying close to the constricted section.
 \begin{figure}[htbp]
   \includegraphics[width=0.50\textwidth]{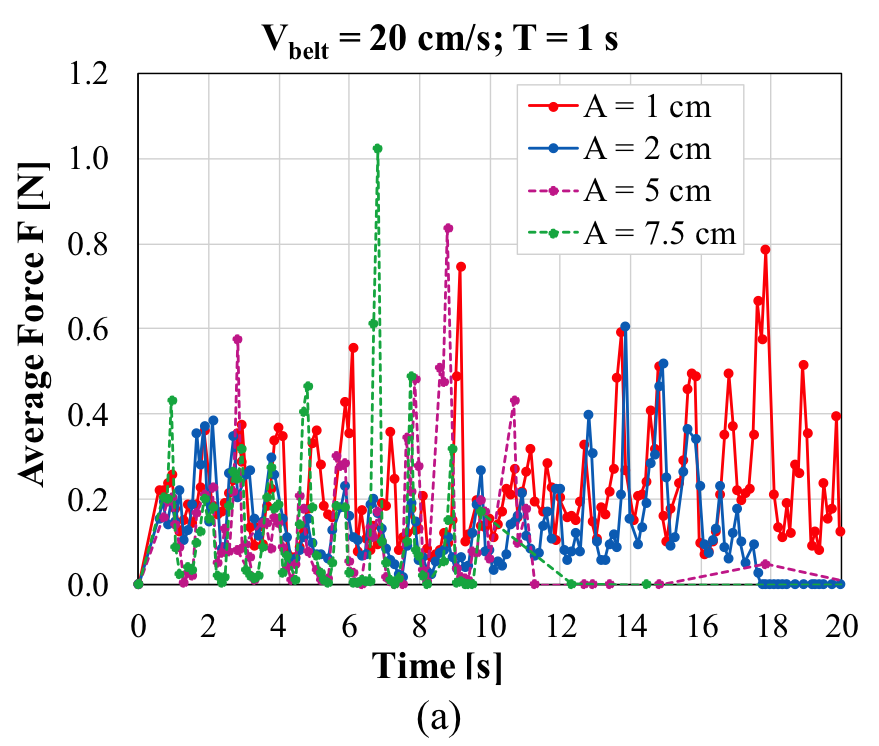}
   \includegraphics[width=0.50\textwidth]{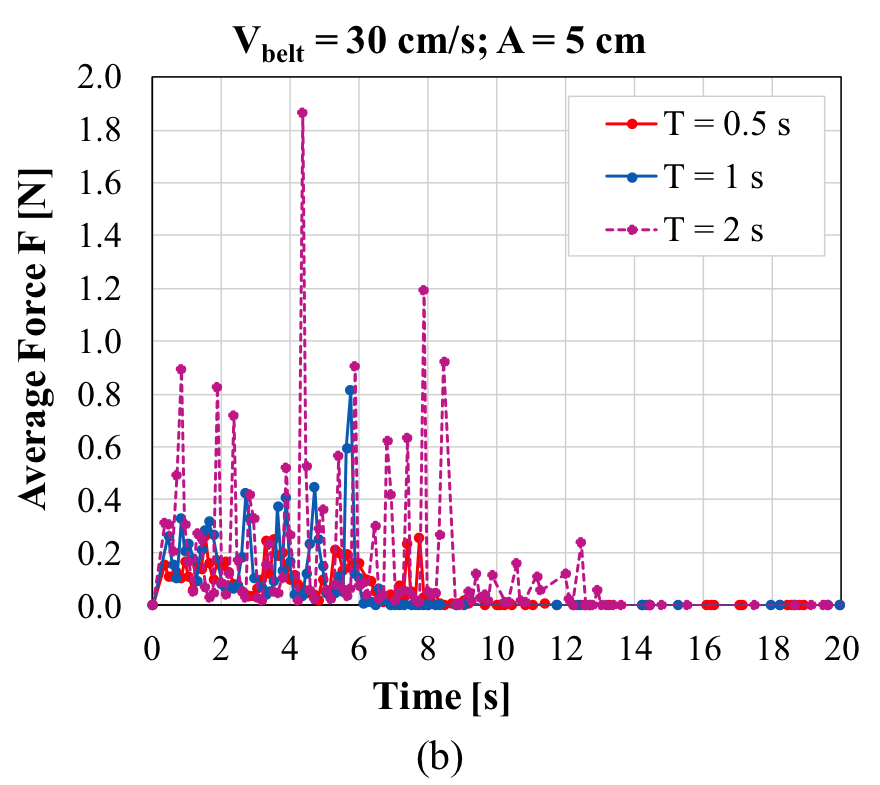} \\
     \includegraphics[width=0.50\textwidth]{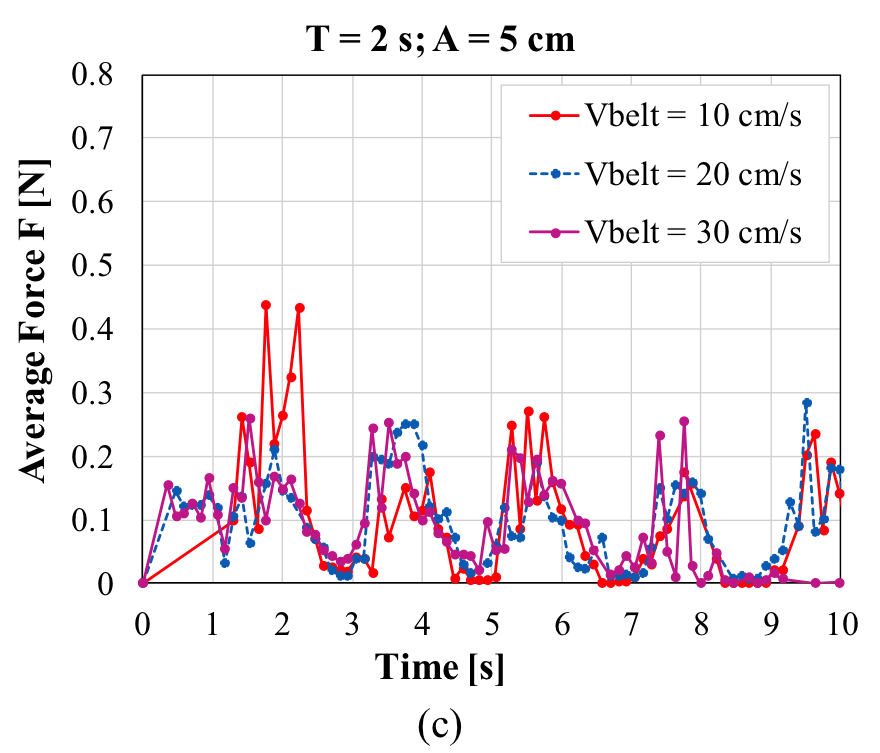}
  \caption{Selection of the optimal amplitude geometry with the force averaged over the first line (Average force) as a function of time for different values of the parameters. (a)  $V_{\text{belt}}=\SI{20}{\cm \cdot sec^{-1}}$, $T=\SI{1}{sec}$ and different amplitudes $A$; (b) $V_{\text{belt}}=\SI{30}{\cm \cdot sec^{-1}}$, $A=\SI{5}{\cm}$ and different periods $T$; (c) $T=\SI{2}{sec}$, $A=\SI{5}{\cm}$  and different values of $V_{\text{belt}}$.
  \label{fig:fig5}}
\end{figure}
We explored four different values of the amplitude $(A = 1, 2, 5, \SI{7.5}{\cm})$, three different values of the belt velocity ($V_{\text{belt}} = 10, 20, \SI{30}{\cm \cdot \sec^{-1}})$, and three different periods $(T = 0.5, 1, \SI{2}{\sec})$ aiming at finding the configuration minimizing the average value of forces acting on the spheres. Here 'average' means average over all considered spheres (coinciding with the force acting on that sphere in the case of a single sphere). The evolution of these forces was then monitored for all tagged spheres shown in Figure \ref{fig:fig4}. Figure \ref{fig:fig5}(a) reports the result of the average force versus time for different values of the amplitude $A$ ranging from $\SI{1}{\cm}$ to $\SI{7.5}{\cm}$ for the 'first line' spheres. The smallest possible case ($A = \SI{1}{\cm}$) shows a significant increase of the average force at longer times presumably due to a clogging effect, since the spheres have diameters of about $\SI{1.6}{\cm}$. On the other hand, large values of the amplitude ($A = 5; \SI{7.5}{\cm}$) give rise to very high spikes in the average force. This suggests the choice $A = \SI{2}{\cm}$ as optimal for the amplitude. Similar arguments, hinging upon Figures \ref{fig:fig5}(b) and \ref{fig:fig5}(c) lead to the final choice of $T = \SI{2}{\sec}$, and $V_{\text{belt}} = \SI{20}{\cm \cdot \sec^{-1}}$. As mentioned, all forces in Figures \ref{fig:fig5} were monitored in the first line spheres, but we have also computed the forces acting on all other tagged spheres with similar results.
Interestingly, our final choice of $V_{\text{belt}}$ turns out to be lower than the value $\SI{30}{\cm \cdot \sec^{-1}}$ used in the real production line. Our result then suggests that a lower value of $V_{\text{belt}}$  would help in preserve the integrity of the cartridges, as one could expect on physical basis. These optimal values will then be used in all following calculations. A movie representing the dynamics under this condition is reported in Supplementary Information (Movie 1).
\subsection{Universality of the coordination number in DEM calculations}
\label{subsec:universality}
\begin{figure}[htbp]
    \includegraphics[width=0.50\textwidth]{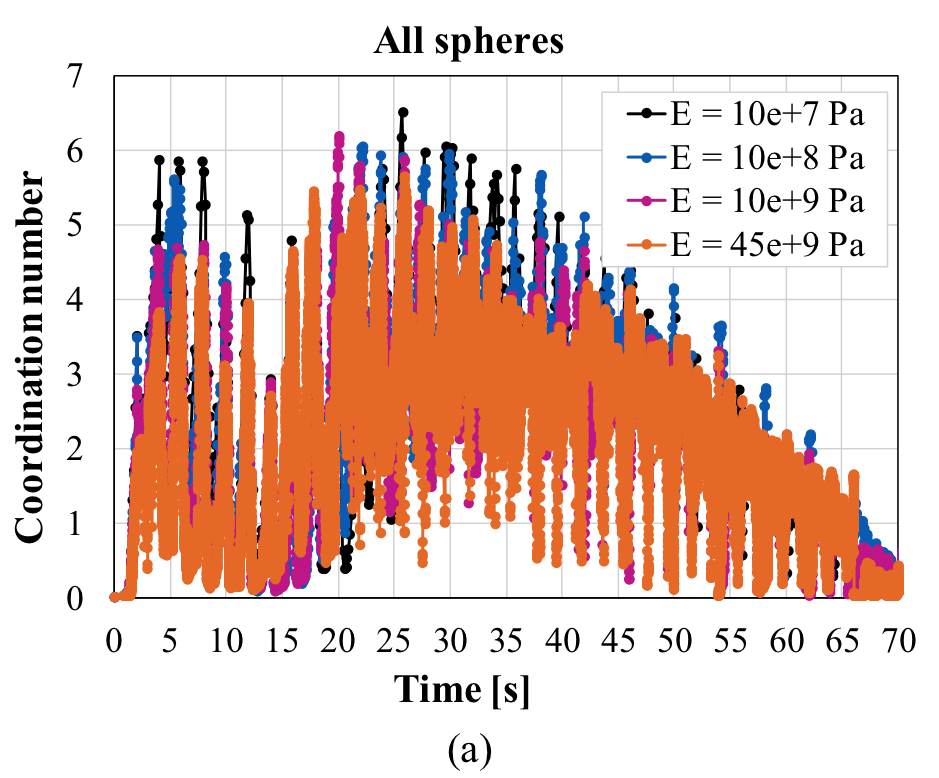}
    \includegraphics[width=0.50\textwidth]{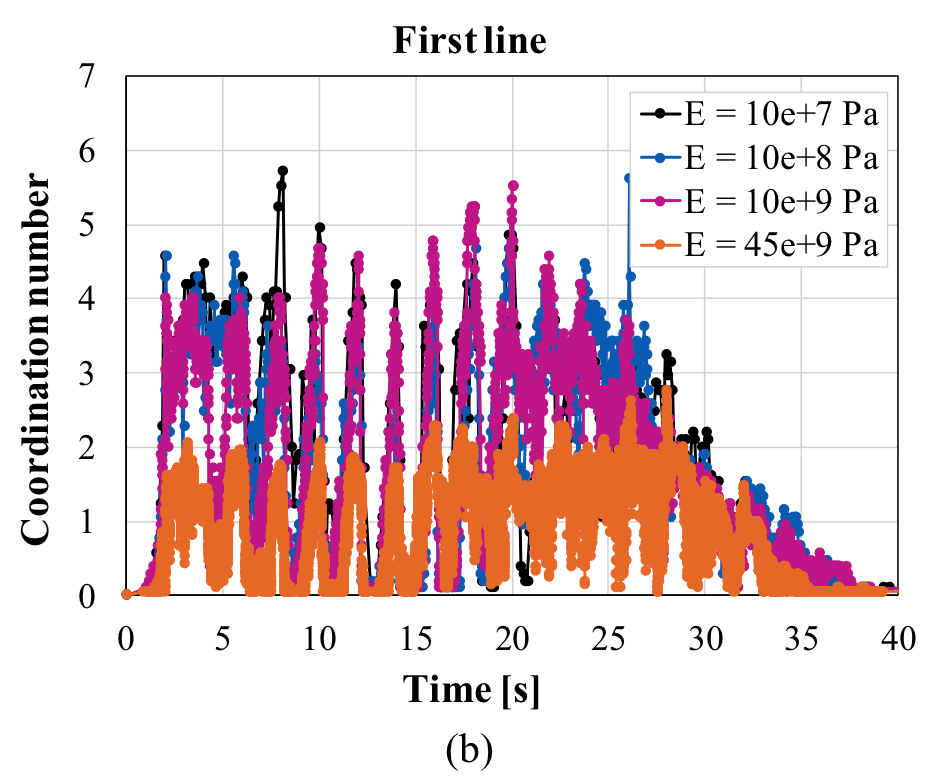} \\
    \includegraphics[width=0.50\textwidth]{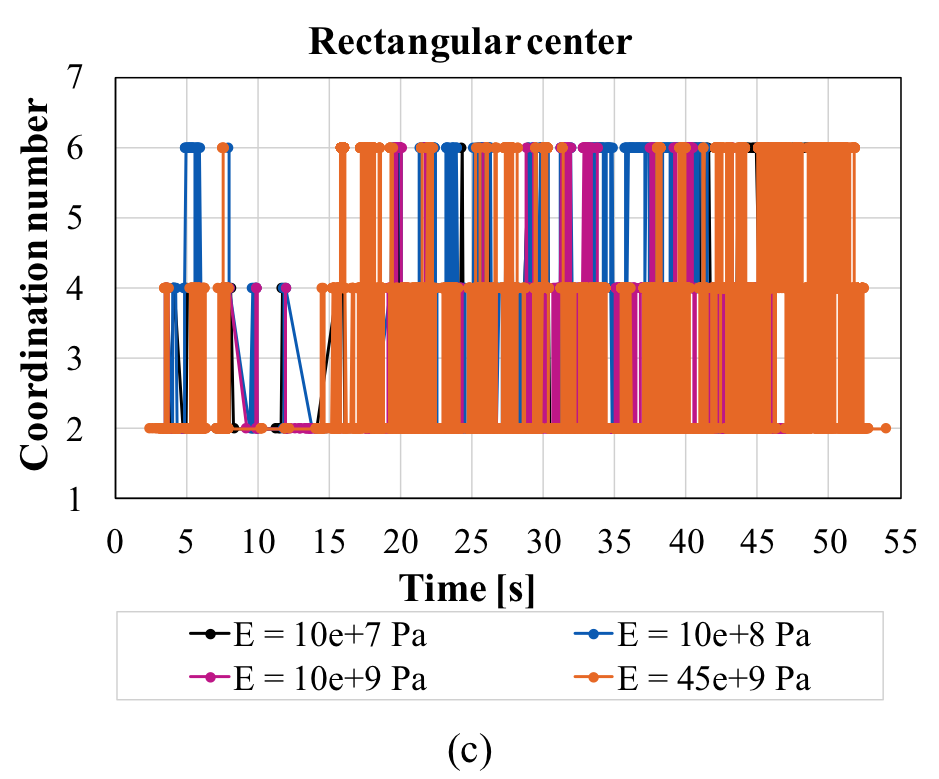}
   \caption{Coordination number as a function of time at increasing values of the Young modulus $E$ for (a) All spheres; (b) 'first line' spheres ; (c) ' rectangular center' sphere. In all cases $V_{\text{belt}}=\SI{10}{\cm \cdot sec^{-1}}$, $T=\SI{2}{sec}$, $A=\SI{2}{\cm}$.
  \label{fig:fig6}}
\end{figure}
In close analogy with granular materials \cite{Poschel10}, we define the 'coordination number' as the number of neighboring spheres in contact with a given sphere. Figure \ref{fig:fig6} reports the results of this calculation as a function of the time for different choices of the tagged spheres and for different values of the Young's modulus $E$. In Figures \ref{fig:fig6}(a) and \ref{fig:fig6}(b) the coordination number is averaged over all spheres and first line spheres, respectively. Figure \ref{fig:fig6}(c) shows the result for the 'rectangular center' sphere that clearly displays an alternations between the two packing structures. Note that the initial configuration is a square packing, as visible from the snapshot of the initial conformation (see Figure \ref{fig:fig4}).

Two key messages can be learned from this analysis. Firstly, the average coordination number is nearly independent of the Young's modulus $E$, thus suggesting that the way the spheres pack at the accumulation table does not depend on $E$. This means that the coordination number can be safely computed at low $E$, with much smaller computational endeavour. The second point is related to the structural rearrangement signalled by the oscillation between 4 and 6 coordination pertaining to a square or hexagonal packing, respectively.  Note the oscillating period that is identical to the driving oscillation of the opening amplitude ($\SI{2}{\sec}$), as one could expect from the outset.

\subsection{Scaling of the velocity field in DEM calculations}
\label{subsec:scaling}
\begin{figure}[htbp]
   \includegraphics[width=0.60\textwidth]{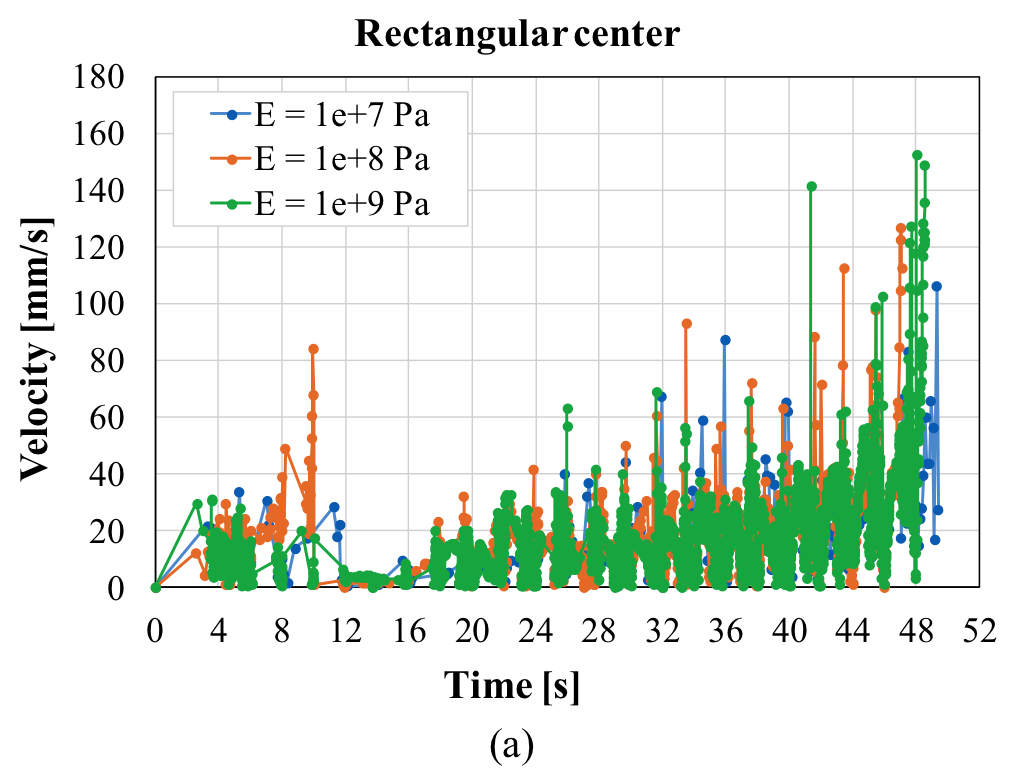}
   \includegraphics[width=0.60\textwidth]{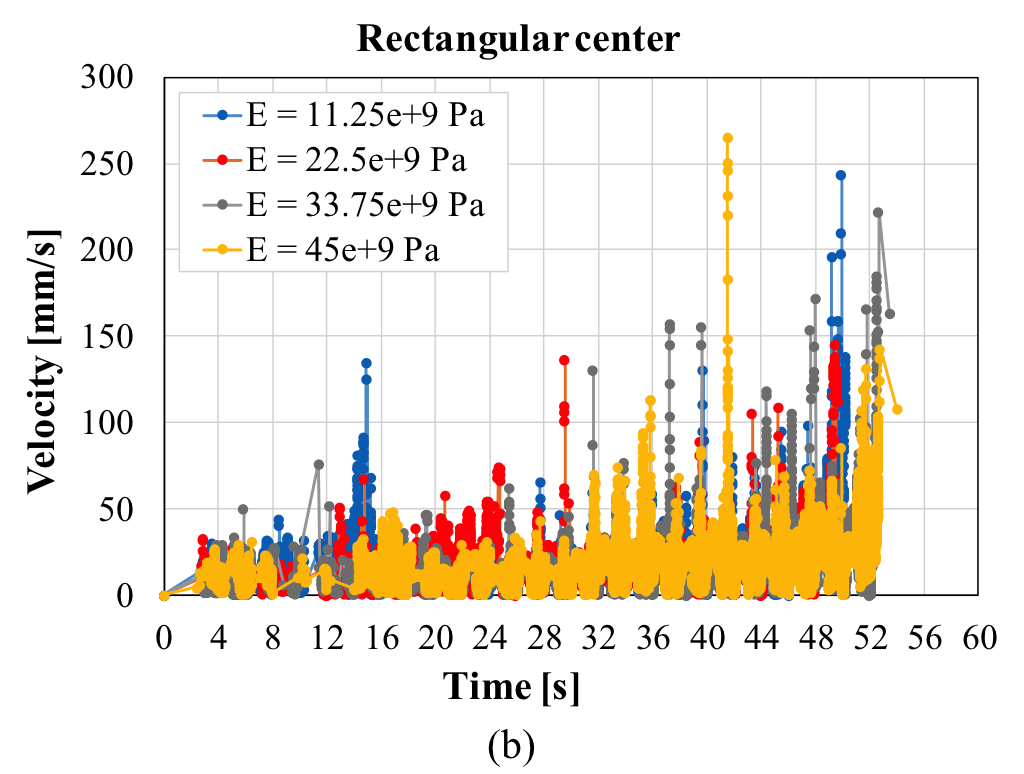}
  \caption{Velocity as a function of time for the ``rectangular center'' sphere  (a) Young modulus range $10^7 \text{Pa}  \le E \le 10^9 \text{Pa}$; (b) Young modulus range $11.25 \times 10^9 \text{Pa}  \le E \le 45 \times 10^9 \text{Pa}$
  \label{fig:fig7}} 
\end{figure}
Unlike the coordination number, contact forces are clearly strongly dependent on the Young's modulus. We can infer this dependence by performing the same calculation - i.e. same initial conditions and same parameters other than the Young's modulus - at different values of the Young's modulus $E$, again as a function of time. Rather than reporting the magnitude of the contact forces, it proves more useful to report the magnitude of the velocity of the spheres. This is because the velocity field will be eventually input to the FEM analysis, so it is important to have a clear picture of its behaviour as a function of time.

We have explicitly monitored the velocities of all tagged spheres. Here we only consider the 'rectangular center' sphere because it best represents the 'bulk behaviour' of the systems. The other values can be found in Figures S3 and S4 of Supplementary Information. In view of the large difference in the values of forces depending on Young's modulus, we have found expedient to gather plots in groups depending on the range of E values. Figure S3a of Supplementary Information reports the results for values $E = 10^7,10^8,10^9\text{Pa}$ whereas Figure S3b of Supplementary Information refers to values of the Young's modulus $E$ from $11.25 \times 10^9 \text{Pa} $ to $45 \times 10^9 \text{Pa}$, the highest of which corresponding to the realistic value in the experimental set-up.

As visible in Figures \ref{fig:fig7}(a) and \ref{fig:fig7}(b), the 'rectangular center' sphere will reach the highest velocities after $\approx \SI{50}{\sec}$, with velocities ranging from $\approx \SI{100}{\mm \cdot \sec^{-1}}$ for the lowest
Young's modulus value of $E=10^7 \text{Pa}$ to  $\approx \SI{250}{\mm \cdot \sec^{-1}}$ for the highest (and realistic) value of $E=45 \times 10^9 \text{Pa}$. These values will be used later on in the FEM analysis.
\subsection{Velocities versus coordination number from DEM calculations}
\label{subsec:contact}
\begin{figure}[htbp]
   \includegraphics[width=0.50\textwidth]{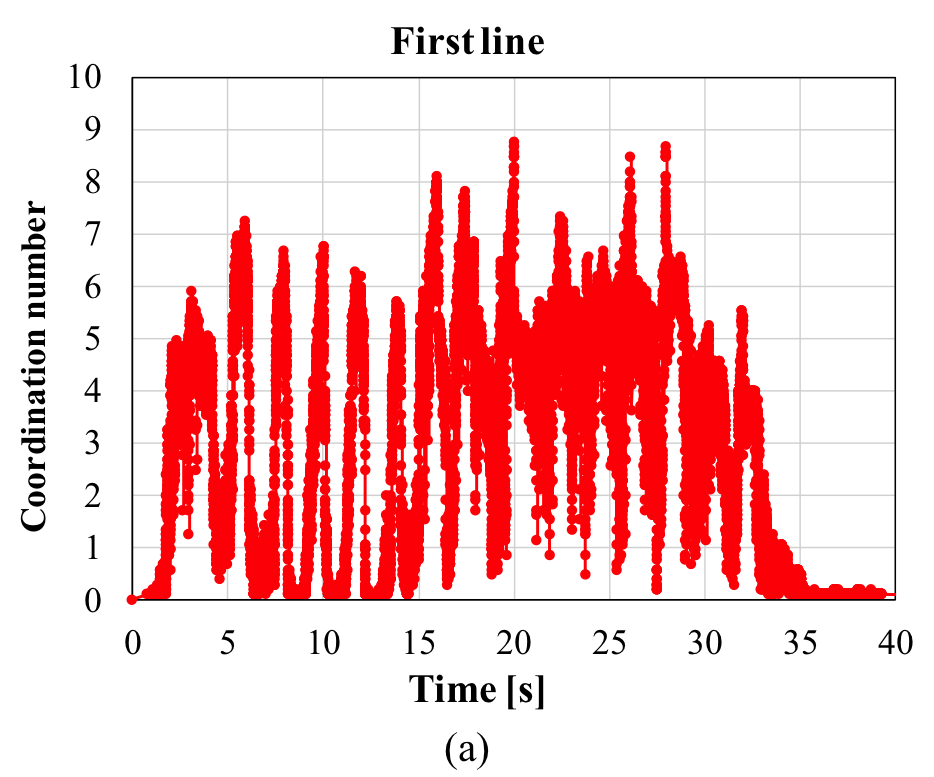}
   \includegraphics[width=0.50\textwidth]{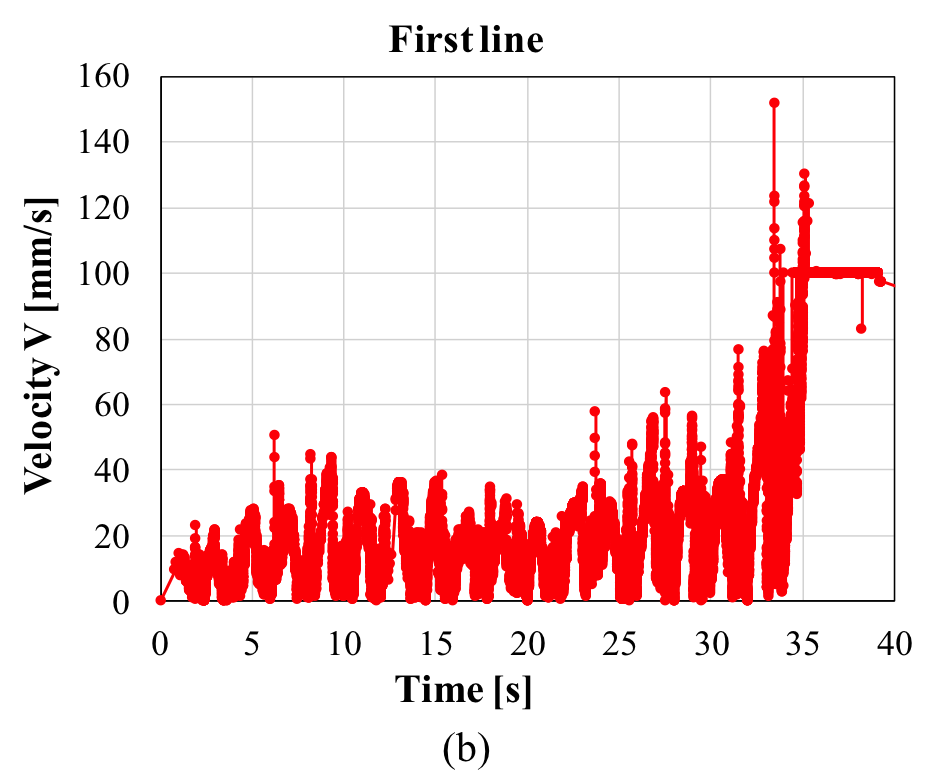} 
 \caption{(a) Coordination number and (b) velocity of the 'triangular center' sphere as a function of time.
  \label{fig:fig8}}
\end{figure}
\begin{figure}[htbp]
   \includegraphics[width=0.40\textwidth]{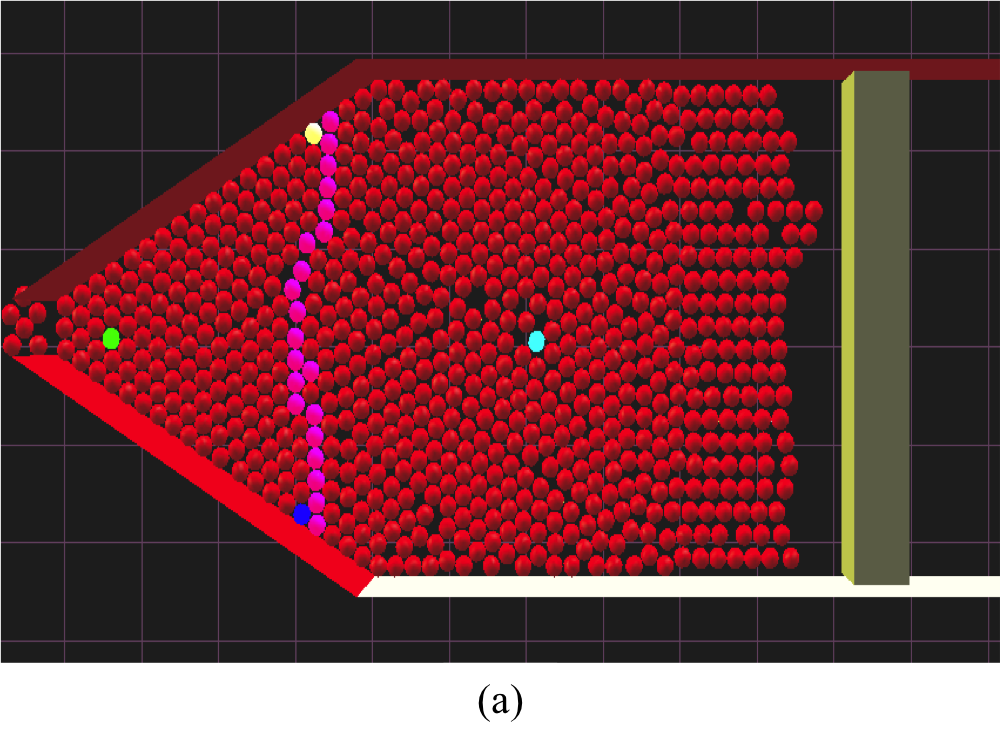}
   \includegraphics[width=0.50\textwidth]{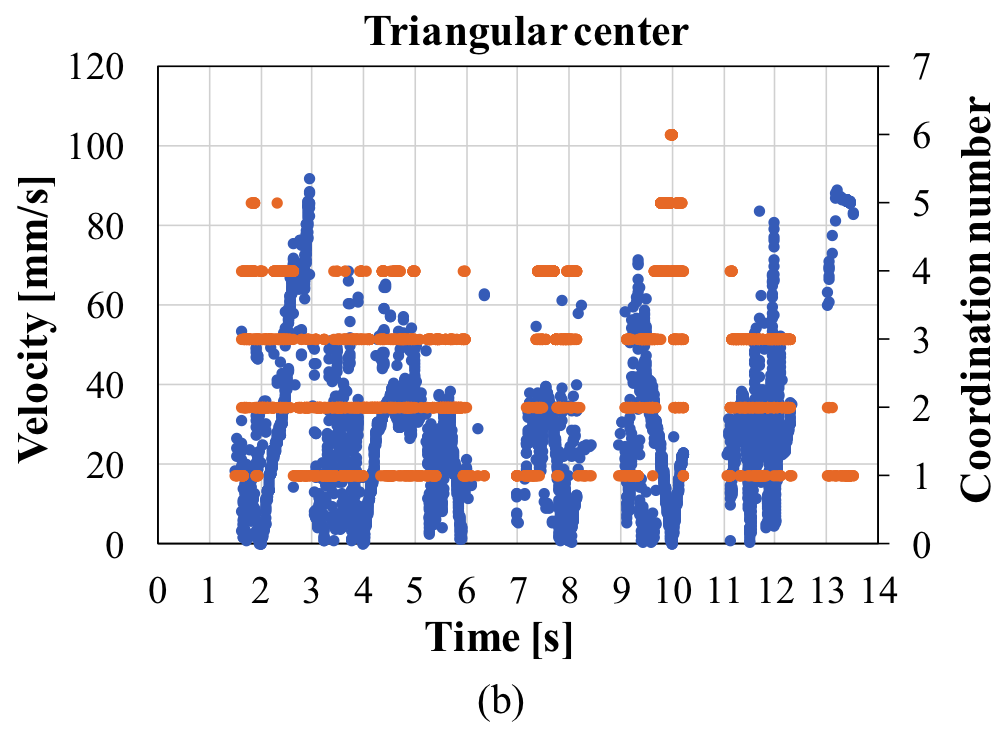} \\
   \includegraphics[width=0.40\textwidth]{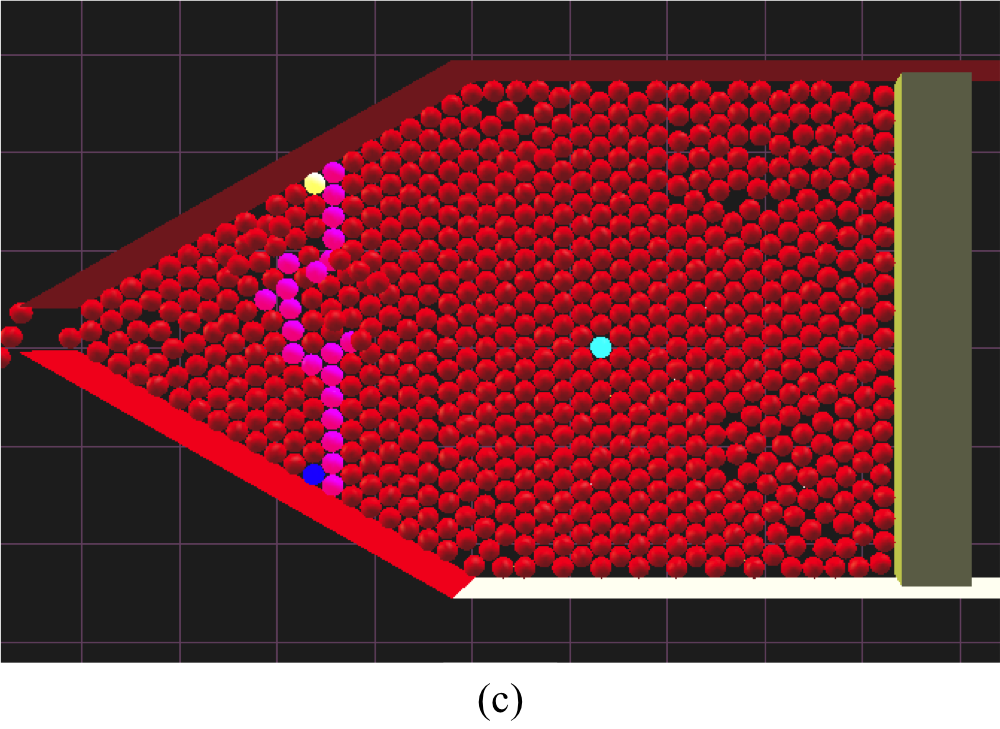}
   \includegraphics[width=0.50\textwidth]{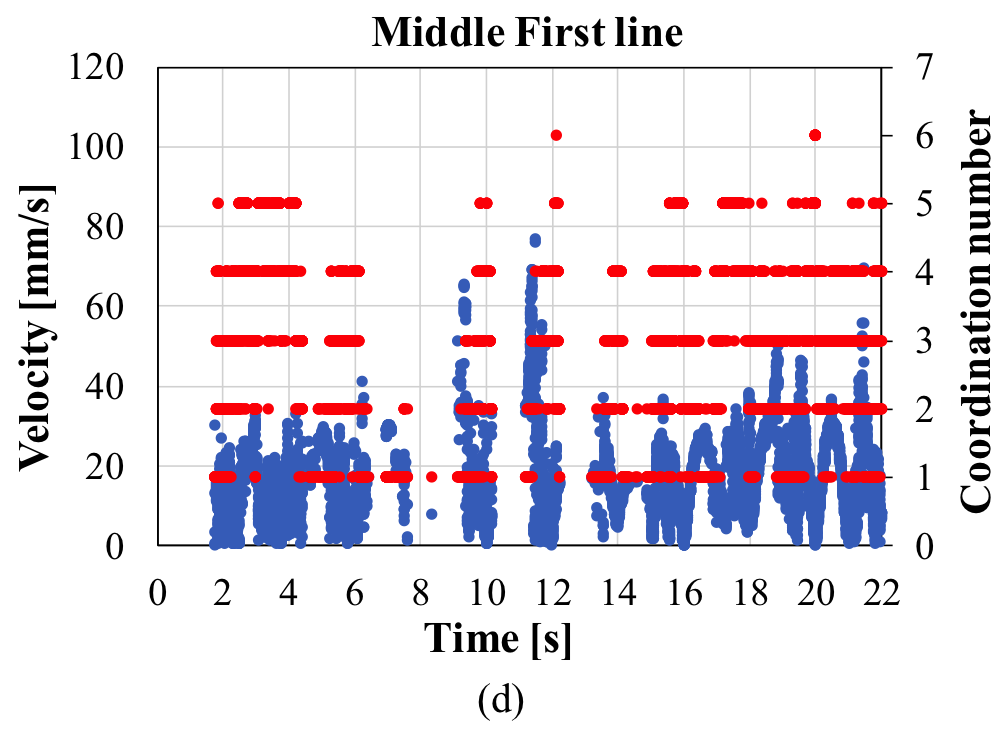}
 \caption{(a) Snapshot of the configuration after $\SI{9.3}{sec}$; (b) Velocity of the 'triangular center' sphere as a function of time;
   (c) Snapshot of the configuration after $\SI{12.2}{sec}$; (d) Average velocity of the 'first line'  sphere as a function of time (here 'middle' means that we do not include extreme values in the averages;
   In both (b) and (d) the right vertical axis also reports the corresponding values of the coordination numbers.
  \label{fig:fig9}}
\end{figure}

Another important outcome of the DEM analysis that will be exploited in the following FEM calculation is associated with the two following questions. How is the velocity field distributed among the different spheres? And is the resulting velocity deriving from single or multiple contacts of the spheres?

We now address these issues by contrasting the resulting velocities with the corresponding coordination numbers for the highest value $E = 45 \times 10^9 \text{Pa}$ of the Young's modulus, since this is the actual value expected in the accumulation table. As a preliminary calculation, we performed a full span of the entire dynamics of all spheres, as well as of the 'first line' spheres. This is done in Figure \ref{fig:fig8} that displays the coordination number (a) and the corresponding average velocities (b) for the 'triangular center' spheres. This is a rather useful information as it allows to associate the velocity of a given group of spheres with the corresponding coordination numbers at the same time. As before, the characteristic periodicity of both quantities is associated with the period of $\SI{2}{\sec}$ of the amplitude oscillation, with the coordination number relaxing from a high value $\approx 6$ corresponding to hexagonal packing, to a nearly vanishing value when the exiting amplitude $A$ is at its largest extension. The velocity also oscillates accordingly around $\SI{20}{\mm \cdot \sec^{-1}}$ until a marked increase is eventually reached when the 'triangular center' spheres reach the outlet after $\approx \SI{33}{\sec}$.

Having identified the most interesting time intervals of each set of spheres, we then zoomed into them to extract specific patterns. This is done in Figures \ref{fig:fig9}(a) and \ref{fig:fig9}(b) for the 'triangular center' sphere, and in Figures \ref{fig:fig9}(c) and \ref{fig:fig9}(d) for the 'first line' spheres. In first case, the 'triangle center' sphere is the closest one to the outlet (labelled in green), and Figure \ref{fig:fig9}(a) shows a snapshot of the configuration after $\SI{9.3}{\sec}$ of simulation, whereas Figure \ref{fig:fig9}(b) displays the corresponding dynamical evolution of the velocity (left axis) and of the coordination number (right axis). After an initial lag of roughly $\SI{2}{\sec}$, where the sphere is nearly immobile into the initial hexagonal conformation, the sphere starts to move with bursts of activities having $\SI{2}{\sec}$ intervals, with velocities and coordination number oscillating  from $\approx \SI{80}{\mm \cdot \sec^{-1}}$ and $5-6$ respectively, to nearly vanishing values. The two quantities appear to be out of phase, however, as one could expect on a physical ground, with the highest velocity achieved when the coordination number is low and conversely lowest velocities associated with high coordination. After approximately $\SI{12}{\sec}$ the 'triangular center'  sphere exits from the amplitude and its dynamic is no longer of interest. This is also visible from the snapshot of the simulation (Figure \ref{fig:fig9}(c)) after $\SI{12.2}{\sec}$ where the (green) 'triangular center' sphere is no longer visible. In this case, the interest switches to the 'first lines' spheres, and Figure \ref{fig:fig9}(d) reports the corresponding average velocity and coordination number. This set of spheres is particularly interesting because it was initially at the boundary of the constricted section with square packing of spheres on the rectangular part of the computational box, and hexagonal packing on the triangular side. In this case, there is an initial slow dynamics with maximum velocities of $\SI{30}{\mm \cdot \sec^{-1}}$ and coordination numbers fluctuating from $1$ to $5$. Here it is important to notice that the tagged spheres that were originally aligned become shuffled in the sea of all other spheres already after $\SI{9}{\sec}$, as it can be clearly seen from the snapshot of Figure \ref{fig:fig9}(a) (taken after $\SI{9.2}{\sec}$ of simulations).

\subsection{The case of the cartridges from FEM analysis}
\label{subsec:cartridge}
By examining the dynamics of the sphere population obtained by means of the discrete element method, it was possible to detect the typical scenario in terms of positions, relative velocities, and coordination numbers. These results allow the setting up of a detailed Finite Element Model calculation that will provide a complete stress map of the cartridges, and will then allow to provide a risk assessment for the single cartridge.

All FEM calculation have been carried out using the Abaqus Code Package \cite{Abaqus}. The cartridge mesh is made of three-dimensional linear tetrahedral elements with three degrees of freedom per node, and the base belt is meshed by using rigid elements. This choice has been dictated by the specific geometry of the cartridge that would required a too dense mesh to describe the neck and the bottom of the cartridge with hexahedrons. We have explicitly tested in few specific cases, that this choice is sufficient for our purposes and that it is not necessary to use quadratic or higher order tetrahedrons.  A friction coefficient of $0.2$ is considered for the contact among cartridges, that means that the contact force has both a normal and a tangential component.

\begin{figure}[htbp]
   \includegraphics[width=0.40\textwidth]{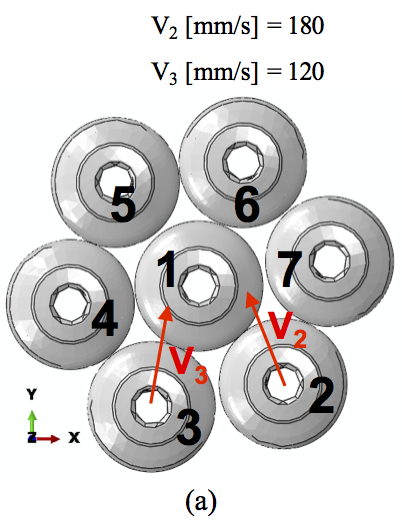}
   \includegraphics[width=0.40\textwidth]{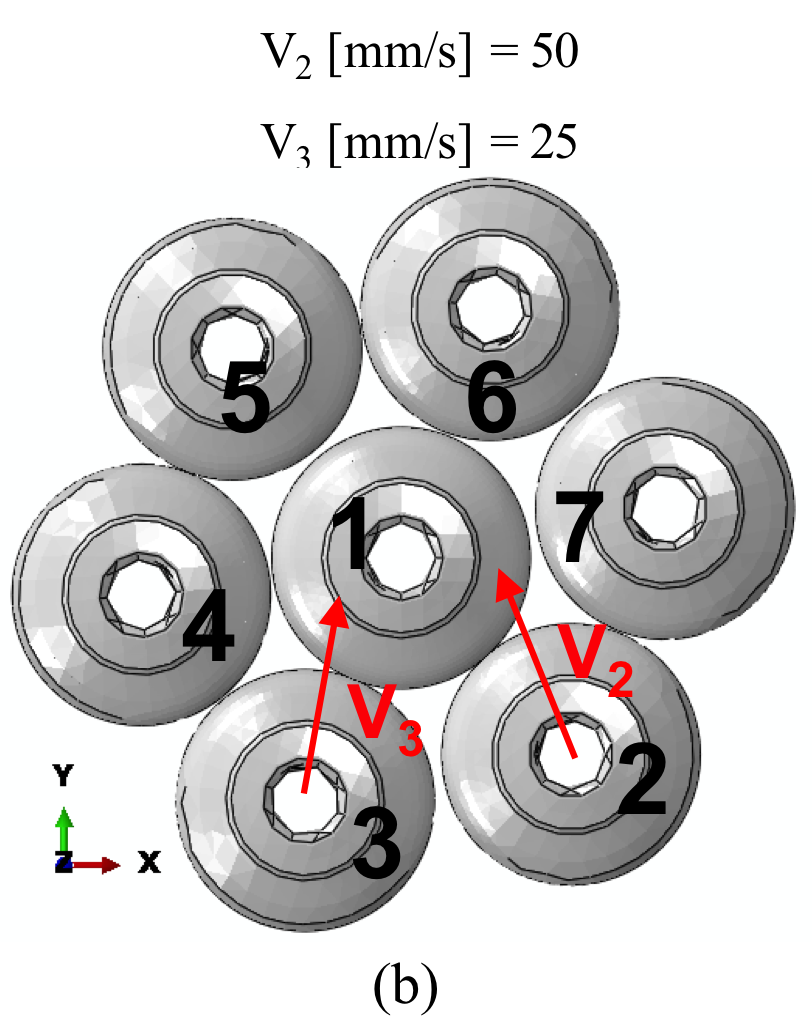}\\
   \includegraphics[width=0.40\textwidth]{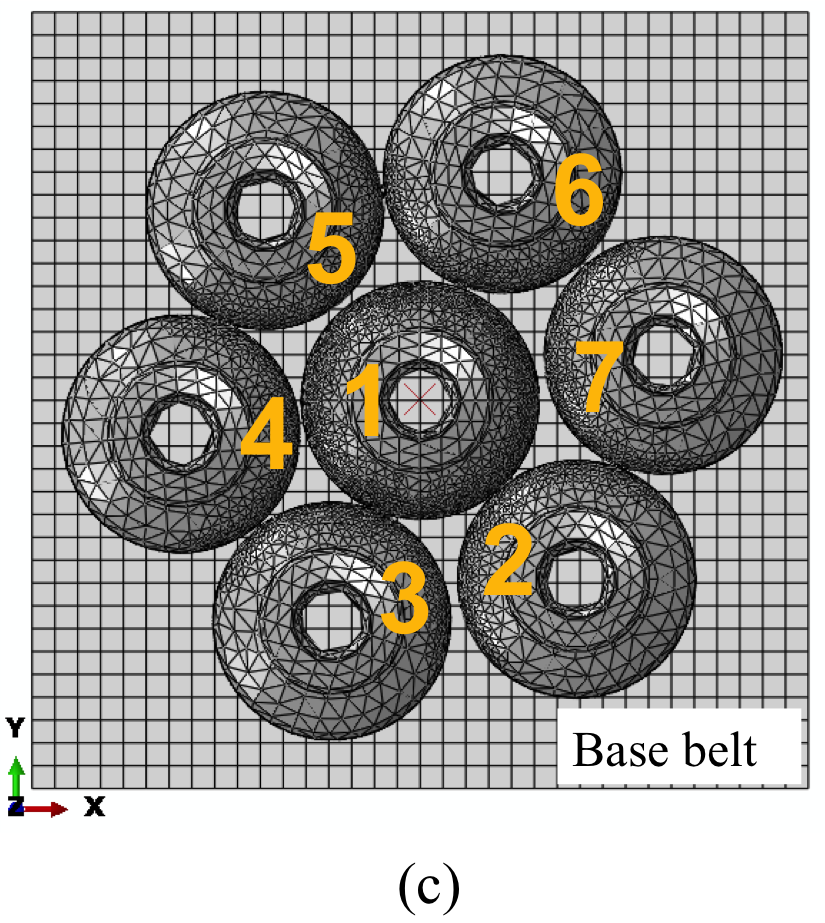}
   \includegraphics[width=0.47\textwidth]{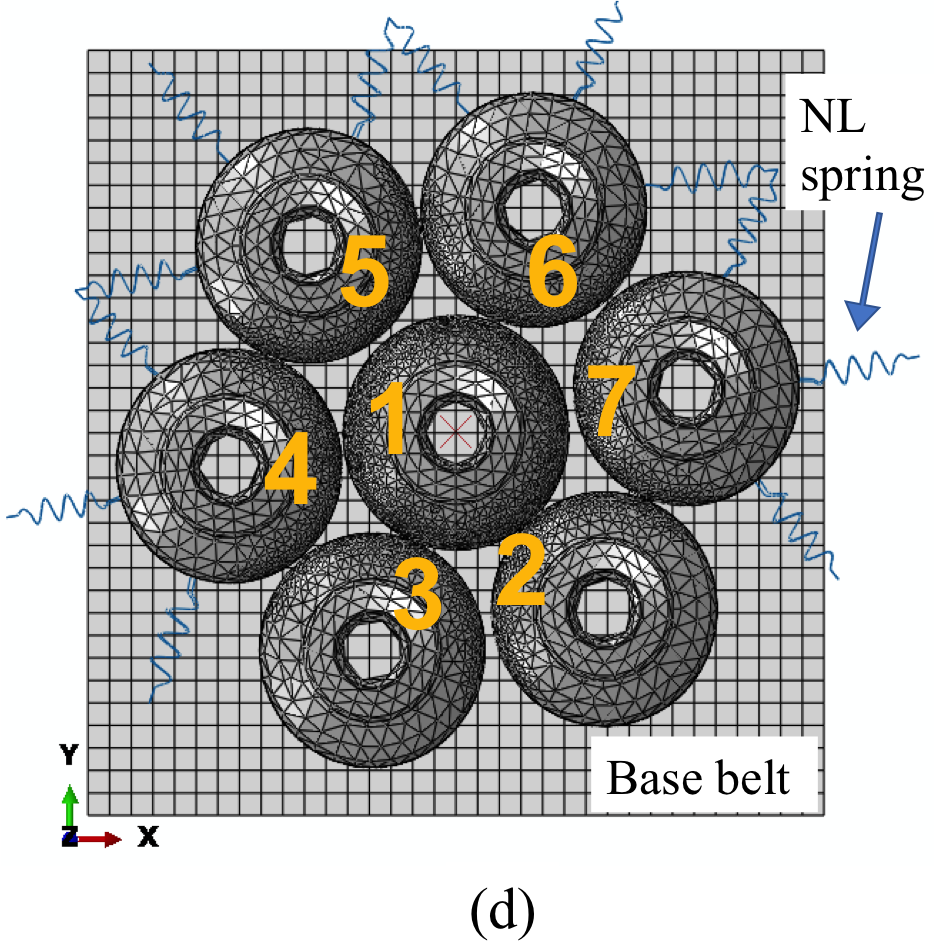}
  \caption{Cartridge positions in the following cases (a) Maximum relative velocity; (b) Average relative velocity; (c) Case of no springs; (d) Case where no-tension springs are considered.
	\label{fig:fig10}}
\end{figure}
\section{Discussion of FEM results}
\label{sec:discussion}
\begin{figure}[htbp]
   \includegraphics[width=0.40\textwidth]{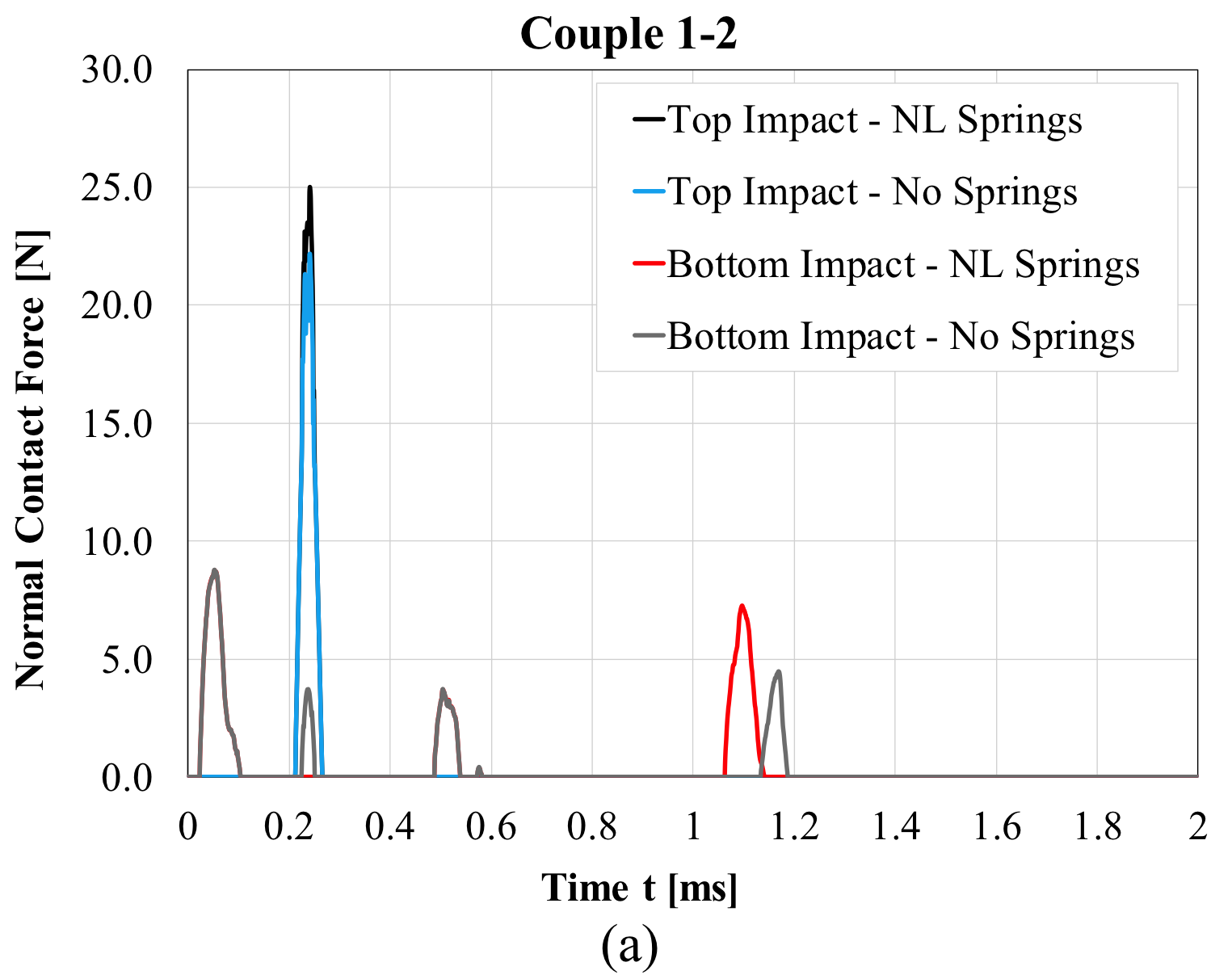}
   \includegraphics[width=0.40\textwidth]{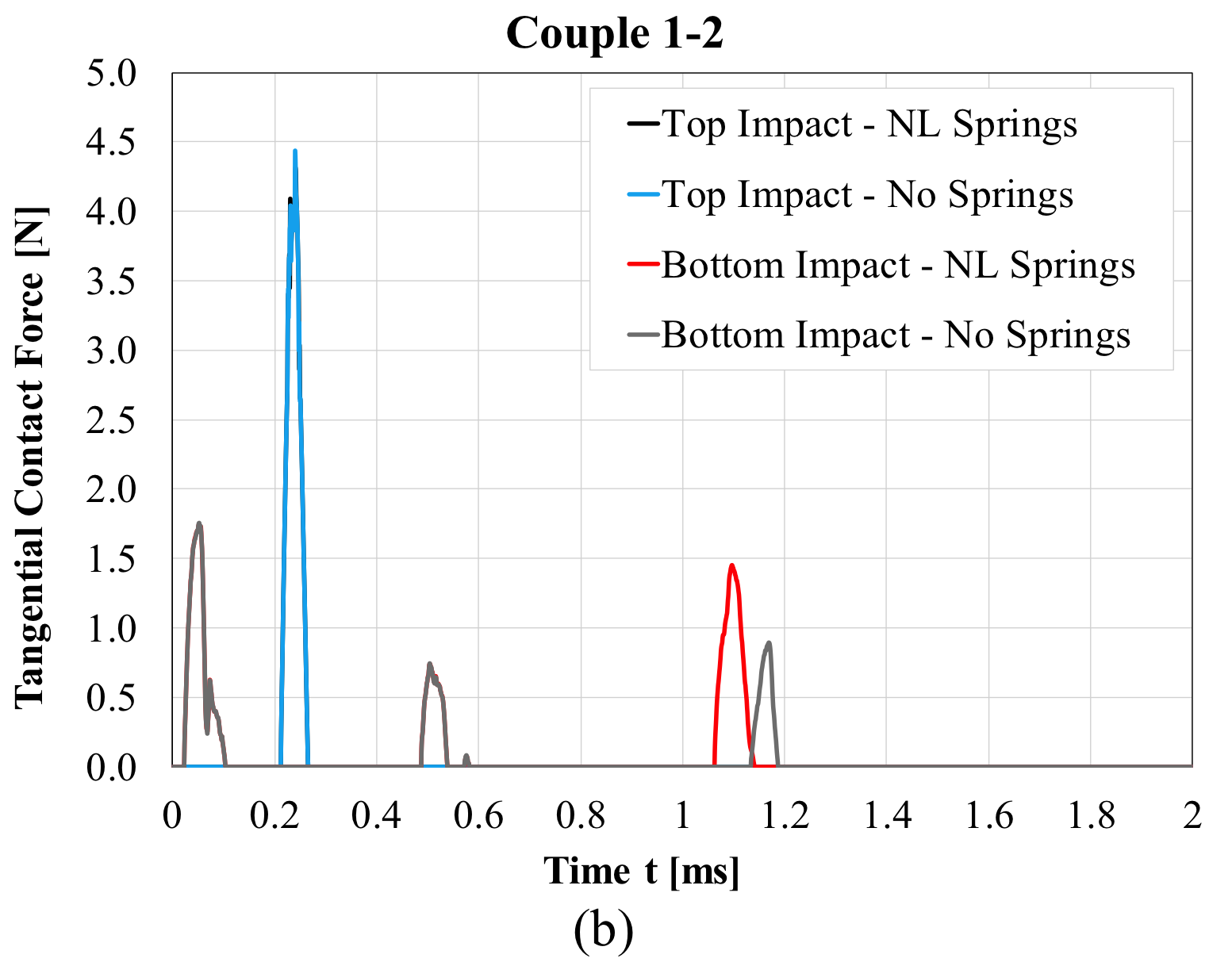}\\
   \includegraphics[width=0.40\textwidth]{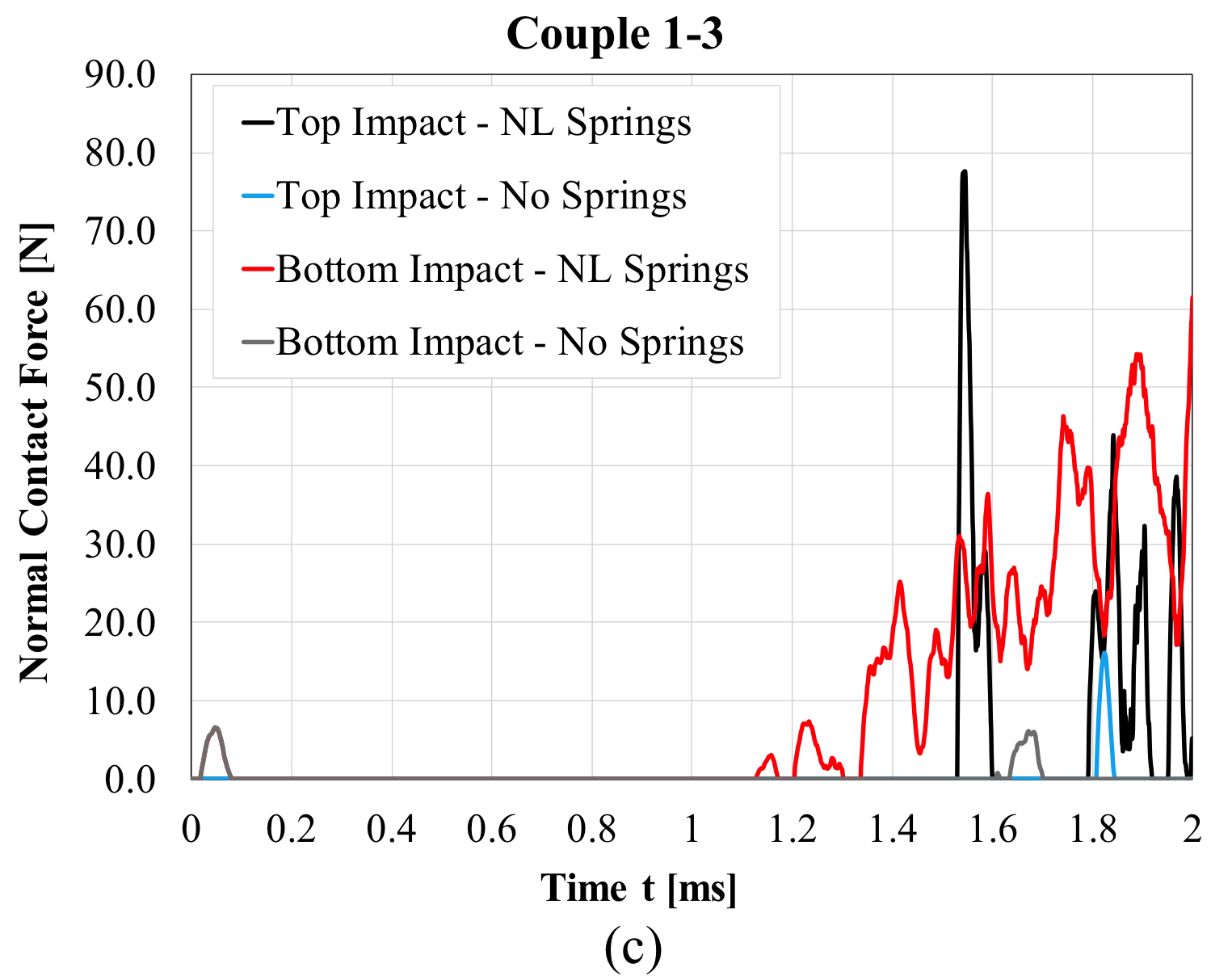}
   \includegraphics[width=0.40\textwidth]{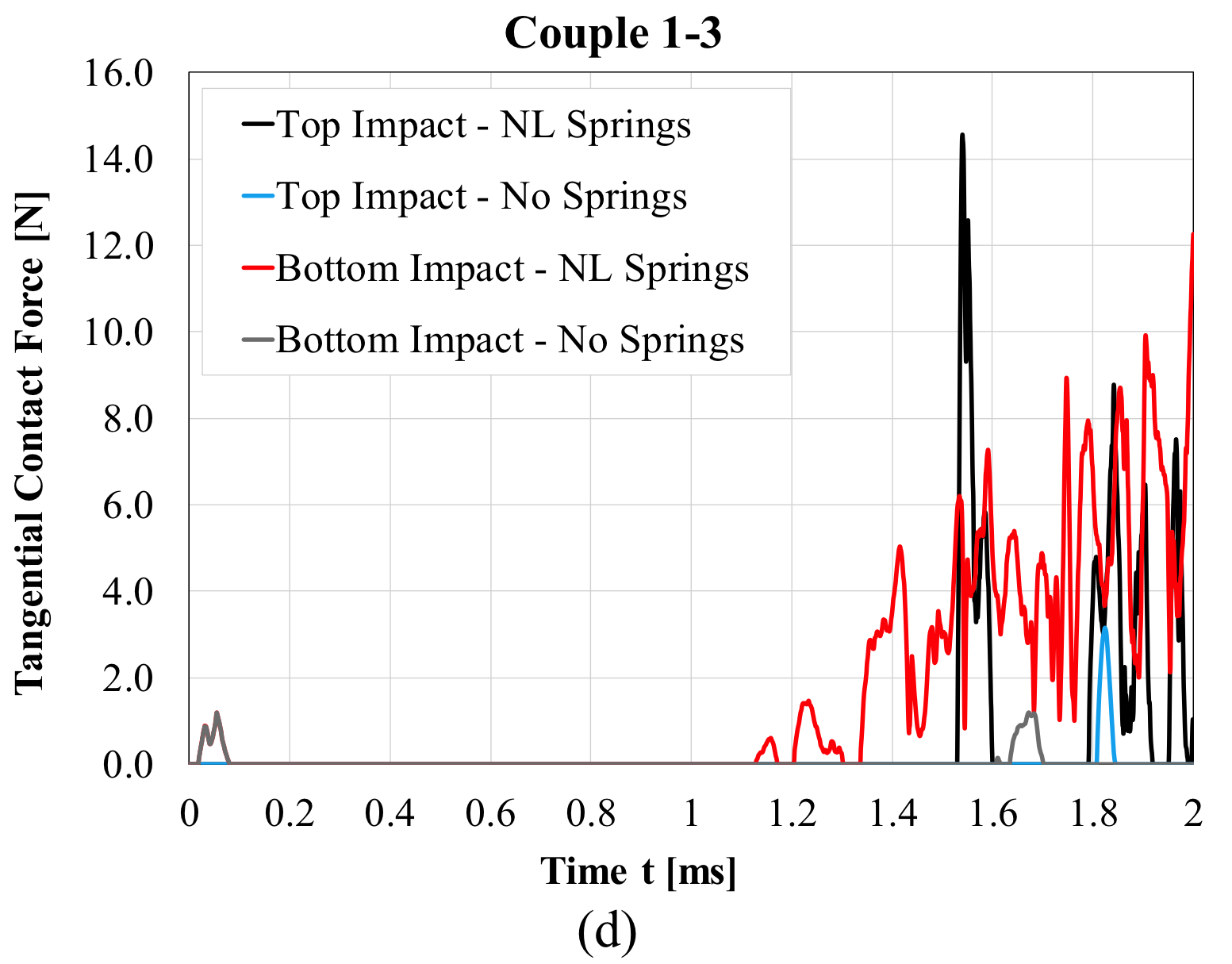}
	\caption{(a) Normal contact force vs time for pair 1-2; (b) Tangential contact force for pair 1-2; (c) Normal contact force vs time for pair 1-3; (d)  Tangential contact force for pair 1-3.
	\label{fig:fig11}}
\end{figure}
Consider a single cartridge (labelled as 1 in Figure \ref{fig:fig10}) surrounded by additional 6 other cartridges (labelled as 2-7 in Figure \ref{fig:fig10}) arranged into a hexagonal conformation that, as we saw from DEM calculation, is the typical conformation where each cartridge is expected to experience the highest stress. We consider the characteristic scenario where two neighbouring cartridges (labelled as 2 and 3) collide with 1 with relative velocities $V_2$ and $V_3$ along a specific direction, and the remaining neighbouring cartridges (labelled 4-7) are in contact with 1.  Based on previously reported DEM analysis, this configuration is one of the most representative among the innumerable possible situations that could take place on the accumulation table. We have then considered the two different cases of \textit{maximum} relative velocities (Figure \ref{fig:fig10}(a)) where $V_2=\SI{180}{\mm \cdot \sec^{-1}}$ and $V_3=\SI{120}{\mm \cdot \sec^{-1}}$), and the case of \textit{average} relative velocity (Figure \ref{fig:fig10}(b)) where $V_2=\SI{50}{\mm \cdot \sec^{-1}}$ and $V_3=\SI{25}{\mm \cdot \sec^{-1}}$) where, in both cases, the colliding angles have been obtained by the velocity vectors of the DEM analysis at selected positions.  For each of the two above cases we have further considered two different conditions. In the first one, all the 7 considered cartridges are uncoupled from the others (Figure \ref{fig:fig10}c)), whereas in the second one cartridges 4-7 are coupled through massless springs to neighbouring 'ghost' cartridges (Figure \ref{fig:fig10}(d)). This latter case is meant to implicitly capture the non-linear effects provided by the set of remaining cartridges at the simplest possible level.   
Hence, in the first case the confinement provided by the adjacent cartridges is neglected (corresponding to a group of cartridges slightly isolated from the rest of the cartridges), whereas in the second case the external cartridges (4 to 7) are restrained by using no-tension springs mimicking the confinement provided by the additional cartridges. While we have performed the analysis for both the maximum (Figure \ref{fig:fig10}a)) and the average (Figure \ref{fig:fig10}b))cases, only the most interesting case relative to the maximum velocity will be reported in the following.

Our interest hinges on the evaluation of the contact forces to estimate the risk of rupture for the cartridges under the worst possible conditions. Both the normal and the tangential contact forces are presented, highlighting their dependency on the initial position, relative velocity and mechanical properties of the cartridges. We focus our attention on pairs 1-2 and 1-3 that are those experiencing the collisions. Figure \ref{fig:fig11}(a) shows the normal and tangential components of the contact forces for the 1-2 and 1-3 pairs as a function of time, starting from a conformation where both pairs are not in contact. Results are reported for both the case with and without external springs and forces are monitored on cartridge 1 which collides with cartridges 2 and 3 both at the bottom and at the top.

A marked difference between bottom and top impacts is observed, and the influence of the external springs is more effective on the 1-3 pair. Consider pair 1-2 depicted in Figure \ref{fig:fig11} (a) for the normal and in Figure \ref{fig:fig11}(b) for the tangential components of the experienced force. The bottom impact forces result significantly smaller than the corresponding top forces, both for normal and tangential components. The springs play a role after $\SI{1}{\ms}$, only on the bottom impact: the collision takes place a little earlier with a slightly higher normal and tangential force.
For the 1-3 pair (Figures \ref{fig:fig11} (c) and (d)), during the initial phase of the impact, the springs do not have any influence on the contact, whereas after about $\SI{1.2}{\ms}$ the dynamics of the system dramatically changes. The confinement provided by the outer cartridges increases the contact forces, leading to considerably larger normal forces up to $\SI{55}{\N}$ for the bottom contact and $\SI{80}{\N}$ for the top contact. For tangential forces values as high as $\SI{10}{\N}$ are achieved for bottom contact and $\SI{14}{\N}$ for top contact. 

This effect can be ascribed to a the combination of constraints provided by the base belt, that is driving the whole population, and by the neighboring cartridges (4-7), thus resulting into a more and more blocking in the case of no-tension springs. This phenomenon has been observed repeatedly for several cartridge pairs, revealing that secondary impacts  play a fundamental role. Within the real accumulation table, this situation can be associated to the dangerous case where some elements get stacked into a narrow space while the following are going to impact against them. Movie 2 of Supplementary Information shows the FEM dynamics in the case where external springs are present.
\subsection{Stress, strain field and microfracture}
\label{subsec:stress}
\begin{figure}[htbp]
   \includegraphics[width=0.40\textwidth]{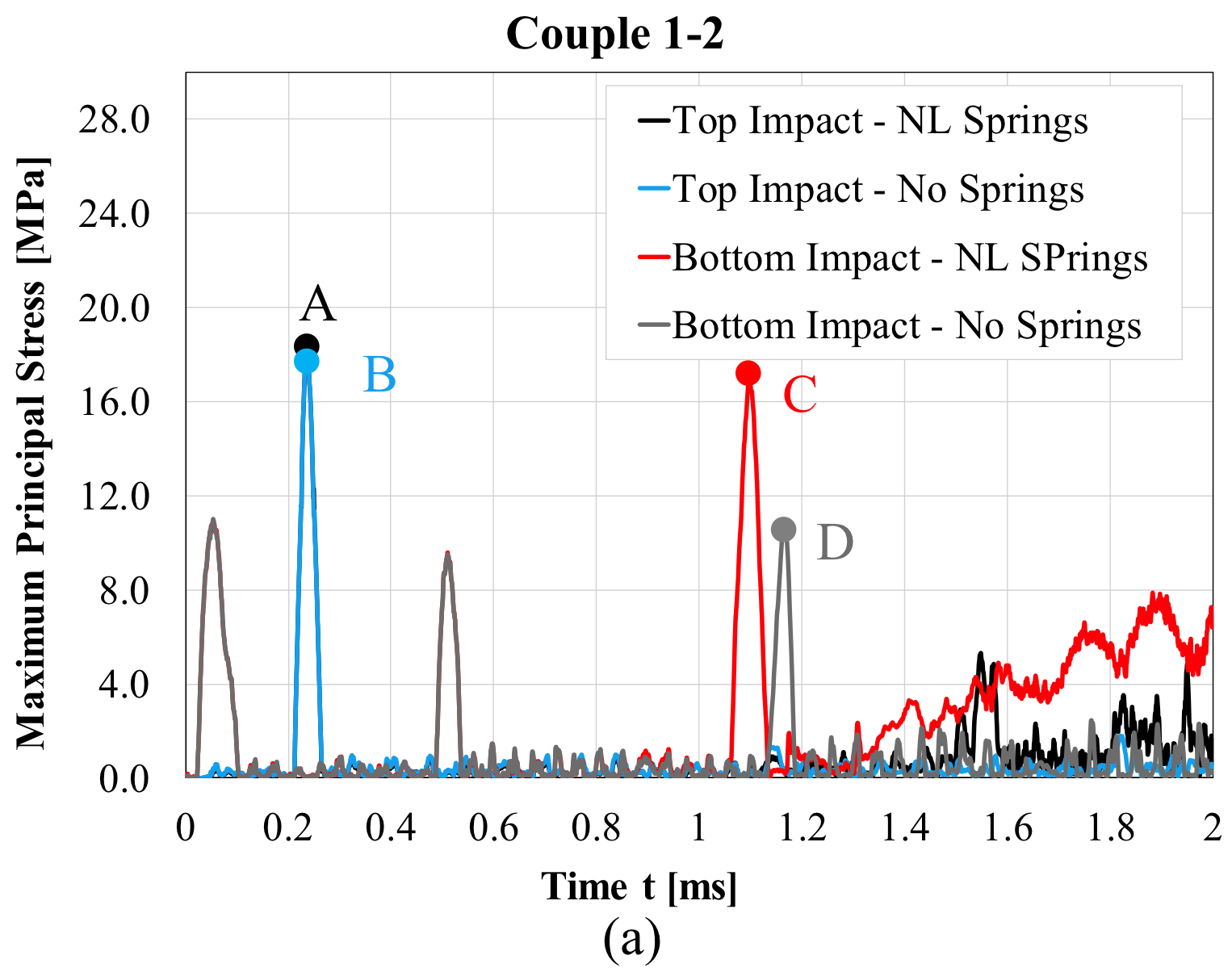}
   \includegraphics[width=0.40\textwidth]{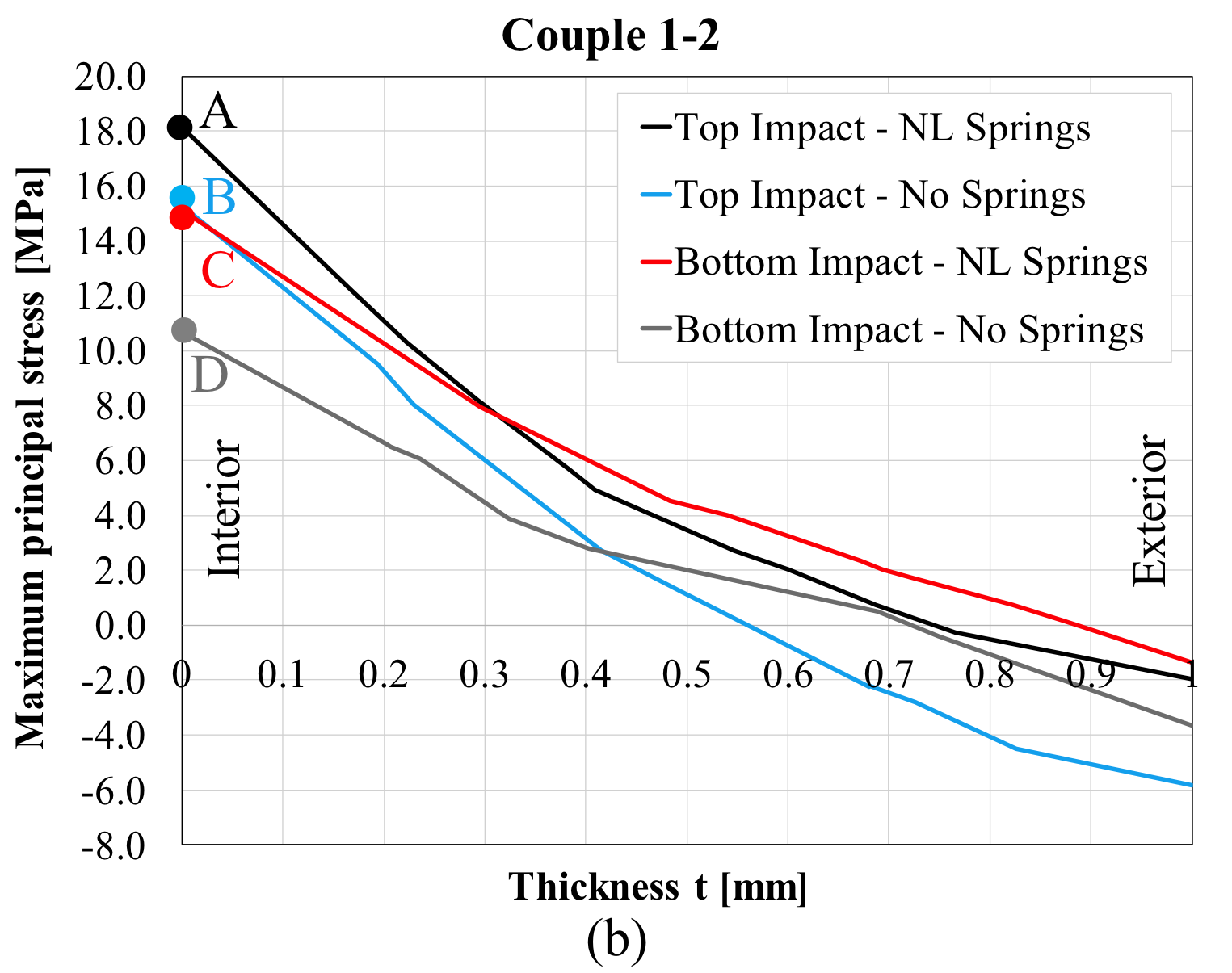}\\
   \includegraphics[width=0.40\textwidth]{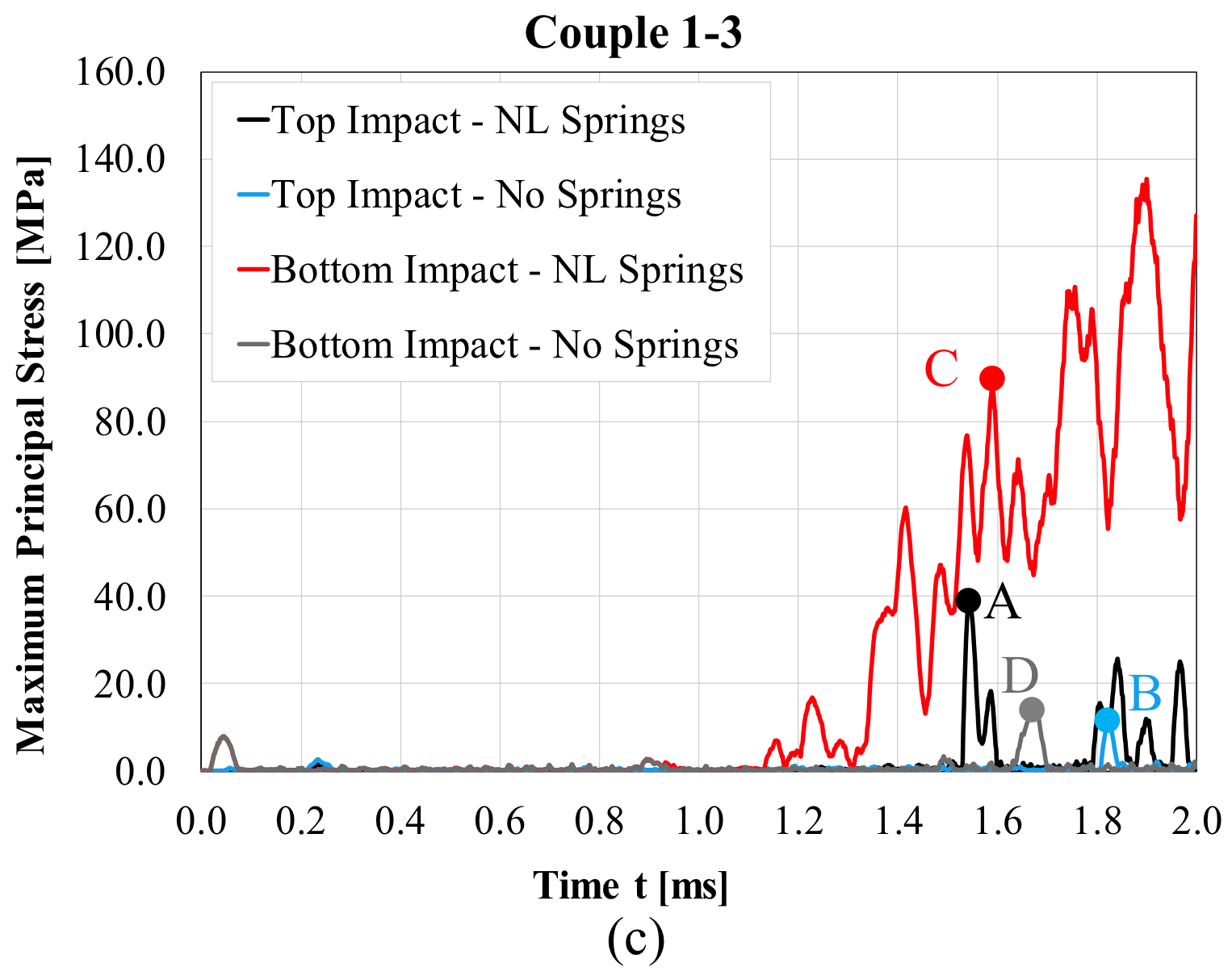}
   \includegraphics[width=0.40\textwidth]{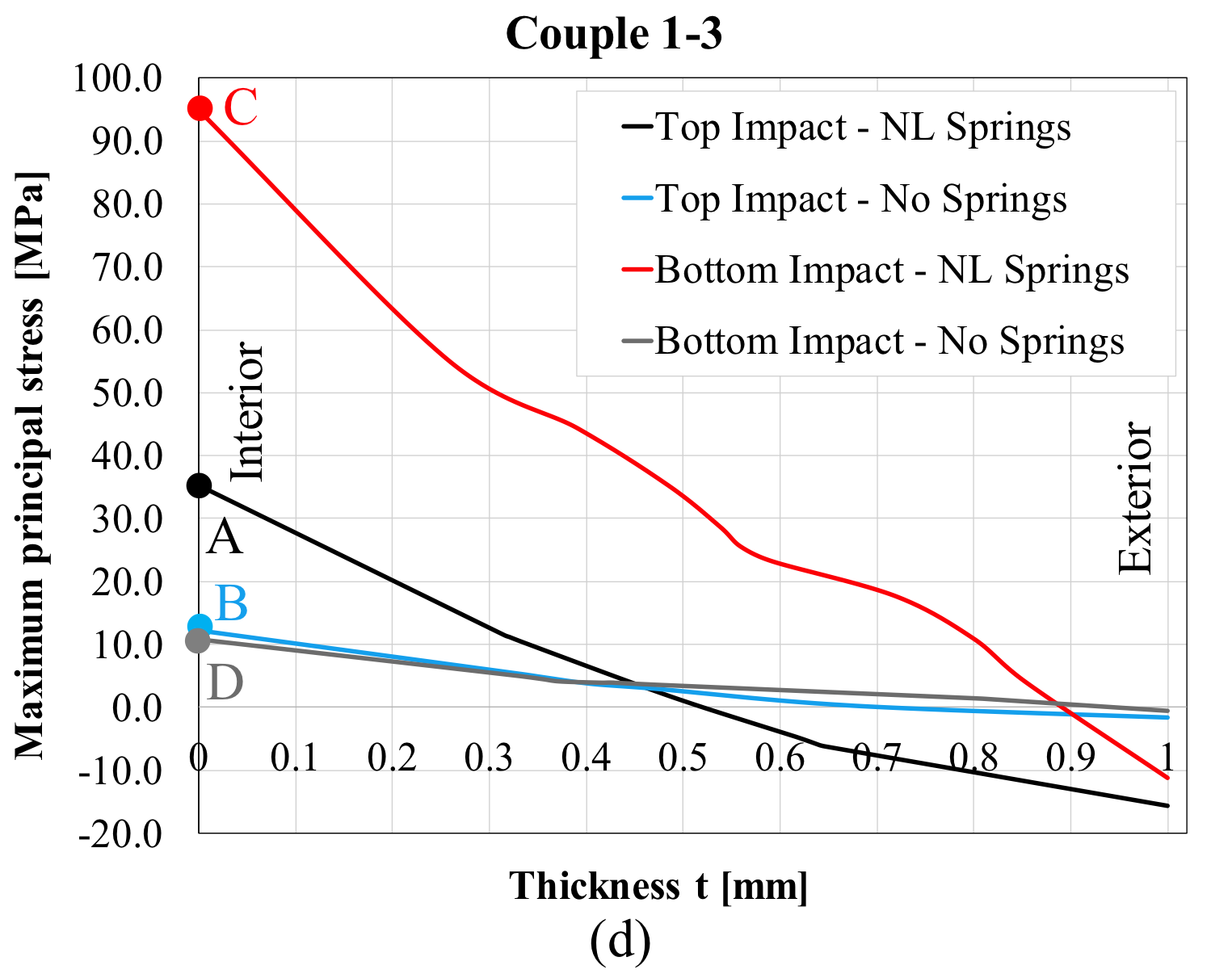}
	\caption{(a) Maximum principal stress vs time for pair 1-2; (b) Maximum principal stress through thickness of cartridge 1 for pair 1-2; (c) Maximum principal stress vs time for pair 1-3; (d) Maximum principal stress through thickness of cartridge 1 for pair 1-3.
	\label{fig:fig12}}
\end{figure}
\begin{figure}
  \centering
  \begin{subfigure}[b]{0.25\textwidth}
    \centering
    \includegraphics[width=\textwidth]{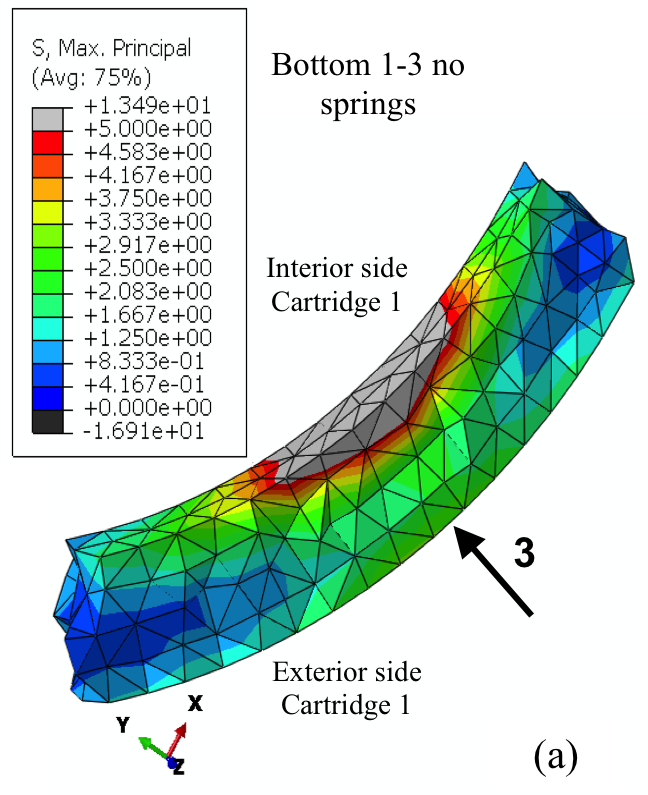}
    \caption{}
  \end{subfigure}
  \hfill
  \begin{subfigure}[b]{0.25\textwidth}
    \centering
    \includegraphics[width=\textwidth]{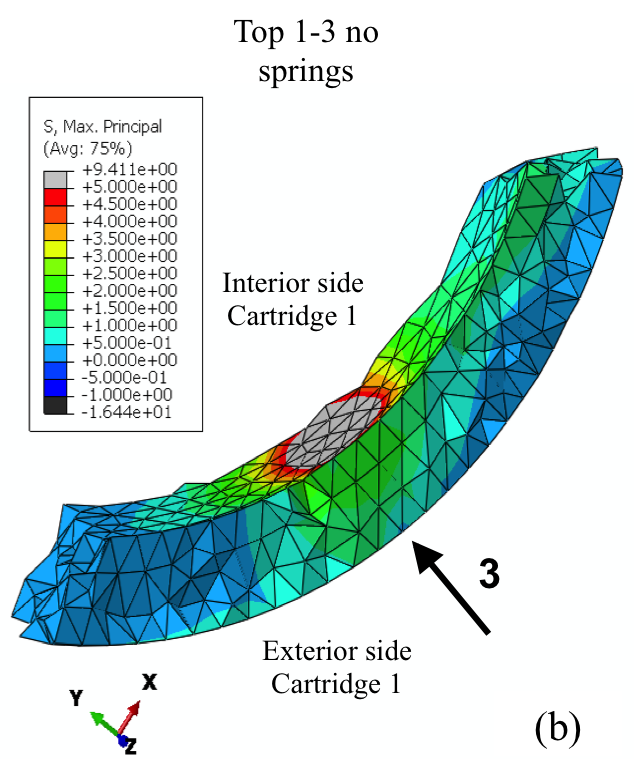}
    \caption{}
  \end{subfigure}
  \hfill \\
  \begin{subfigure}[b]{0.25\textwidth}
    \centering
    \includegraphics[width=\textwidth]{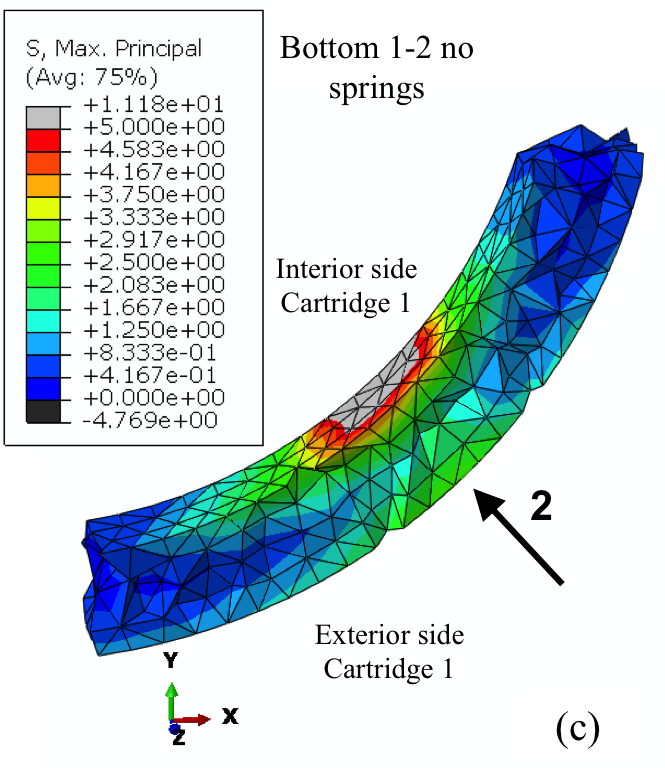}
    \caption{}
    \end{subfigure}
\hfill
  \begin{subfigure}[b]{0.25\textwidth}
    \centering
    \includegraphics[width=\textwidth]{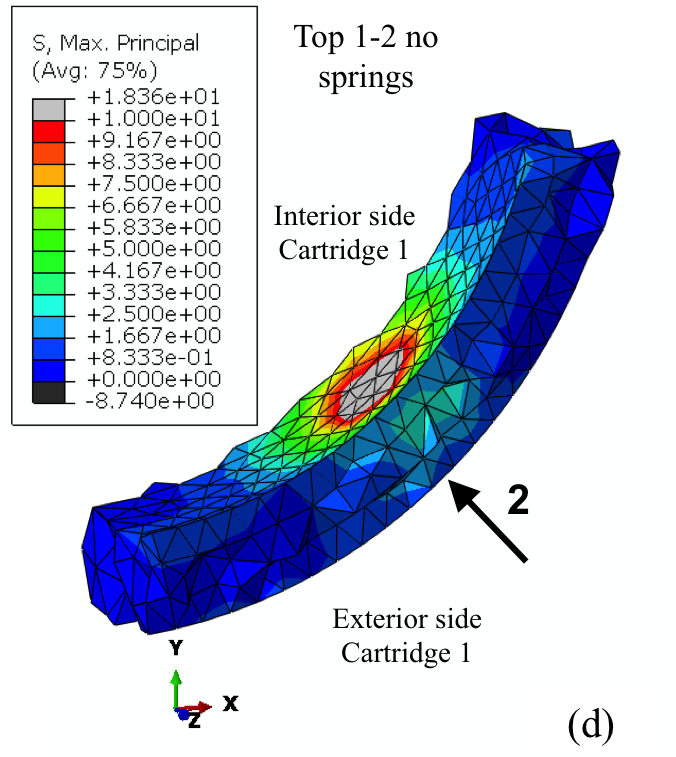}
    \caption{}
  \end{subfigure}
  \hfill \\
  \begin{subfigure}[b]{0.25\textwidth}
    \centering
    \includegraphics[width=\textwidth]{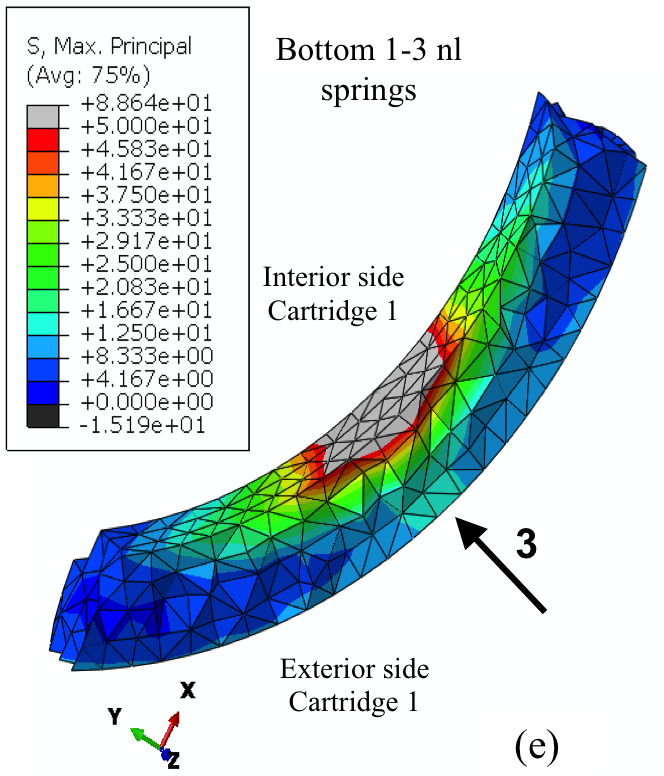}
    \caption{}
    \end{subfigure}
\hfill
  \begin{subfigure}[b]{0.25\textwidth}
    \centering
    \includegraphics[width=\textwidth]{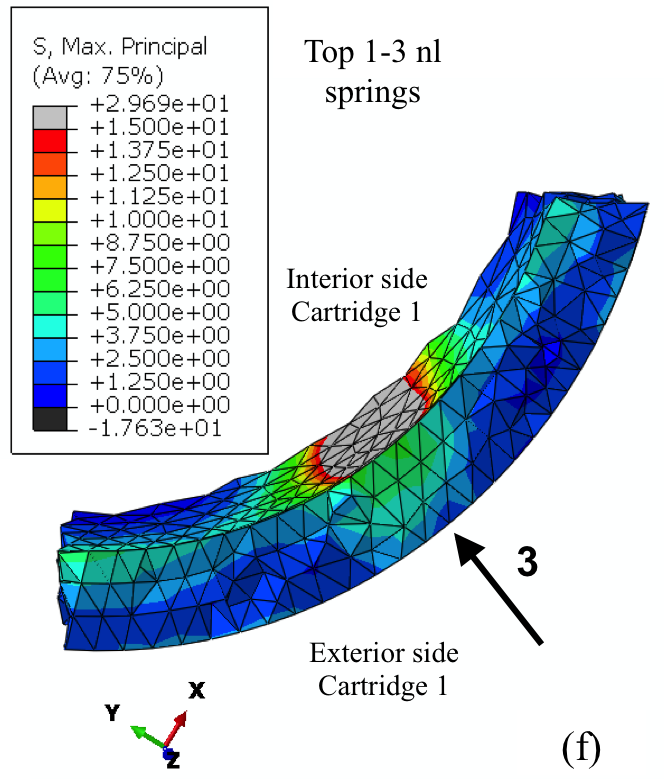}
     \caption{}
  \end{subfigure}
  \hfill \\
  \begin{subfigure}[b]{0.25\textwidth}
    \centering
    \includegraphics[width=\textwidth]{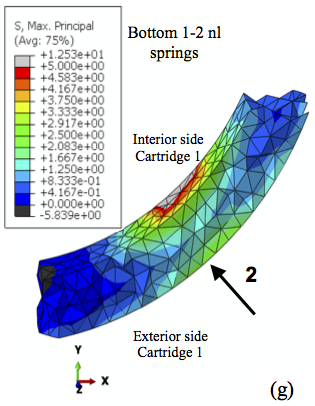}
    \caption{}
  \end{subfigure}
  \hfill
  \begin{subfigure}[b]{0.25\textwidth}
    \centering
    \includegraphics[width=\textwidth]{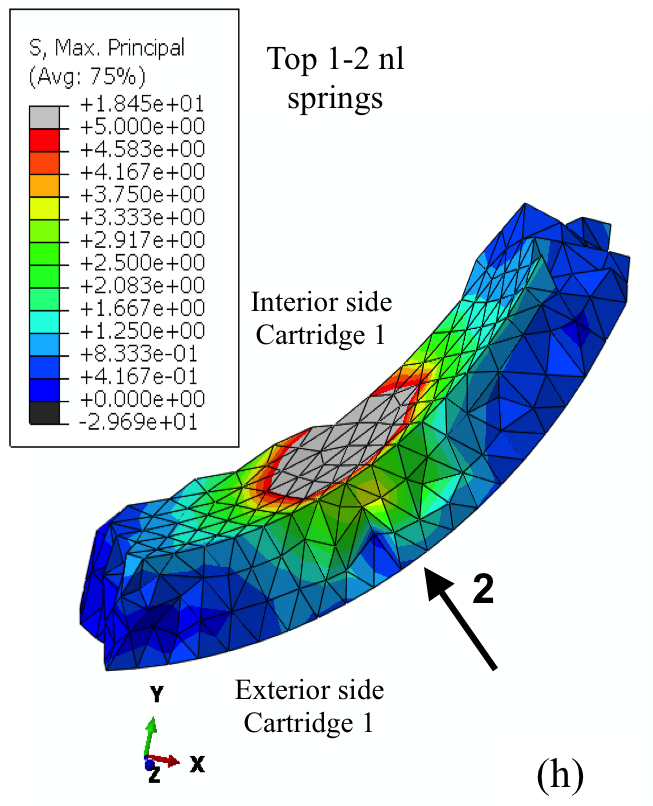}
    \caption{}
  \end{subfigure}
  \hfill \\
   \caption{(a) Pair 1-3 bottom, no springs; (b) Pair 1-3 top, no springs; (c)  Pair 1-2 bottom, no springs; (d)  Pair 1-2 top, no springs;  (e) Pair 1-3 bottom, non-linear springs; (f) Pair 1-3 top, non-linear  springs; (g)  Pair 1-2 bottom, non-linear springs; (h)  Pair 1-2 top, non-linear springs.
	\label{fig:fig13}}
\end{figure}

An essential theme in container integrity preservation is understanding the mechanisms behind the sudden failure and finding possible strategies for the mitigation of its risk. To this aim, a possible risk assessment can be made through the evaluation of the forces and peak stresses developed in the cartridge during its path on the accumulation table. It is very interesting to investigate the variation of the maximum principal stress through the thickness of the glass wall, highlighting its correlation with the impact force. As presented in the Method section, the contact load is directly linked to the relative velocity, the mechanical parameters of the container and in general to the initial conditions. This aspect is exasperated under real conditions on the accumulation table, where the chaotic dynamics of the cartridges causes several, continuously varying shocks. In Figures \ref{fig:fig12}(a) and \ref{fig:fig12}(c) the maximum principal stress is shown as a function of time for pairs 1-2 and 1-3. When considering the confinement springs, for the pair 1-3 the peak stress reaches $\SI{140}{\MPa}$ at the bottom of the cartridge number 1. At the instant of the maximum peak stress, the corresponding variation through the thickness is shown in Figures \ref{fig:fig12}(b) and \ref{fig:fig12}(d). The worst situation is recorded for pair 1-3 with no-tension spring where the maximum principal stress reaches $\SI{94}{\MPa}$ inside the cartridge at the contact zone. It is interesting to observe a shifting trend of the stress, with increasing thickness depth subjected to tension stresses with consequent danger for crack initiation and/or propagation. 
Figure \ref{fig:fig13} reports the contours of the maximum principal stress of cartridges 1 in correspondence of the peaks of normal contact forces for the various cases. The black arrow indicates the impacting cartridge. It is worth to underline that the interior and exterior part of the interested impact zone are in tension and compression respectively. The possible crack initiates from inside the cartridge and propagates toward outside.
This is a very dangerous situation, because the containers can be damaged by microscopic scraps which, in the worst case, can lead to the sudden rupture of the cartridge and cause the stop of the filling line, with high costs for the company.

In terms of risk assessment, by considering a realistic scenario based on DEM kinematics outcomes used as input for the Finite Element modelling, it is possible to extrapolate what could happen in the real accumulation table when the dynamical system is extended to thousands of cartridges.


\section{Conclusions}
\label{sec:conclusions}
The interest in the present paper was triggered by a practical questions with far reaching consequences, but eventually evolved into an interesting general problem on its own right. The filling procedure of glass cartridges used in pharmaceutical applications poses increasing challenges to guarantee the integrity of the cartridge and the patient safety during the injection system. This process was mimicked in our calculation via a sequential combination of DEM-FEM calculation that led to the complete stress field of the cartridges. We used Discrete Element Methods (DEM) to study the full dynamics of a large number of equivalent spheres in an accumulation table of variable geometry. We have motivated the use of equivalent sphere by showing that their dynamics under these conditions is representative of the dynamics of the cartridges (see Movie 3 in SI for a visual representation) that would otherwise be out of the reach of present computational capabilities. In addition to the optimal accumulation table conditions, DEM calculation provided the distribution of the number of contacts, the intensity and directions of the corresponding velocities and positions as function of time within the non-linear elastic collision model. These results were then used as input for the Finite Element Method (FEM) calculation that provided the contact forces as a function of time, the stress tensor field, as well as the critical cracking zones for realistic values of the glass composition. Here, the likelihood of a crack initiation is evaluated based on the maximum principal tensile stress.
It would be extremely interesting to implement a line of experimental investigations where the numerical predictions reported in the present work could be tested in simplified case studies. Among the possible available techniques, a promising example is given by photoelastic force measurements, which have been used with success in experiments with granular materials \cite{Daniels17}. Our findings may pave the way toward a general approach to these class of systems, thus transcending the specific case reported here.


\acknowledgements{The support of a Regione Veneto FSE Grant 2120-38-2216-2016 is gratefully acknowledged.}

%
%

\clearpage

\end{document}